\documentclass[a4, 11pt, reqno, final]{amsart}

\usepackage{amsmath, amscd, amssymb, amsthm, amsfonts}
\usepackage{mathtools, mathrsfs}
\usepackage{graphics, graphicx}
\usepackage{subcaption}
\usepackage[colorlinks=true, urlcolor=blue, citecolor=blue, linkcolor=blue, raiselinks=true, breaklinks=true]{hyperref}
\usepackage{enumerate}
\usepackage{color}
\usepackage[numbers, sort&compress]{natbib}
\usepackage{placeins, caption, subcaption}
\usepackage{tabularx}
\usepackage{arydshln}
\usepackage[colorinlistoftodos]{todonotes}
\usepackage[notref,notcite]{showkeys}
\usepackage[margin=1in, right=4cm]{geometry}
\usepackage[utf8]{inputenc}
\usepackage{tikz, pgfplots, tikz-cd}
\usepackage{adjustbox}
\usepackage[percent]{overpic}

\usepackage{bm}
\usepackage{bbm}

\def\R{\mathbb{R}}

\newcommand{\bfe}{{\bf e}}

\newcommand{\bfu}{{\bf u}}

\theoremstyle{plain}% thmstyleone

\newtheorem{remark}{Remark}

\title{Surface Growth Driven by
an Optimality Criterion}

\author{Rohan Abeyaratne}
\address[R. Abeyaratne]{Department of Mechanical Engineering, Massachusetts Institute of Technology, Cambridge, MA, USA}
\email{rohan@mit.edu}

\author{Roberto Paroni}
\address[R. Paroni]{Department of Civil and Industrial Engineering, University of Pisa, Largo Lucio Lazzarino 1, 56122, Pisa, Italy}
\email{roberto.paroni@unipi.it}

\author{Marco Picchi Scardaoni}
\address[M. Picchi Scardaoni]{Department of Civil and Industrial Engineering, University of Pisa, Largo Lucio Lazzarino 1, 56122, Pisa, Italy}
\email{marco.picchiscardaoni@ing.unipi.it}

\begin{document}

\date{\today}
\begin{abstract}
We propose a variational framework for accretive surface growth driven by an optimality principle. Rather than prescribing a kinetic law, the configuration at each time step is obtained, within a time-discrete setting, as the solution of a constrained minimization problem. Growth is modeled as an irreversible surface deposition process subject to a global mass constraint, while the driving mechanism is encoded in an objective functional, here taken to be the structural mean compliance.

The approach is illustrated on a linearly elastic cantilever beam whose cross-sectional height evolves through layered accretion, possibly involving prestrain and precurvature. Growth-induced residual stresses can alter the convexity of the compliance functional, leading to nonuniqueness and localization phenomena. We explore the possibility of adding a regularization term penalizing deviations from the previous-step configuration. Finally, through a formal limiting procedure, we derive from the time-discrete formulation a time-continuous limit in the form of a constrained gradient flow.
\end{abstract}

\maketitle
\tableofcontents

\section{Introduction}

Surface growth, also known as \emph{accretive growth}, describes the evolution of a body through the addition of material exclusively at its boundary, in contrast to volumetric growth processes in which mass is generated throughout the bulk.
This mechanism governs the development of a wide range of biological and physical systems, including seashells, horns, tree trunks, bones (see, e.g., Thompson~\cite{Thompson1942}), and engineered structures fabricated by additive manufacturing.
In such systems, newly deposited material becomes progressively incorporated into the solid, modifying both geometry and mechanical response over time.

In the mechanics literature, growth is traditionally modeled by prescribing the evolution of internal variables, most commonly a \emph{growth tensor}, through suitable kinetic laws.
A foundational framework was introduced in the seminal work of Skalak et al.~\cite{Skalak1982}, and the general morphoelastic theory of growth was systematically developed following the influential paper of Rodriguez et al.~\cite{Rodriguez1994}; see also the comprehensive monographs by Goriely~\cite{Goriely2017} and Taber~\cite{Taber2020},   {and the special issue \cite{Nardinocchi2025}}.
In this setting, a multiplicative decomposition of the deformation gradient identifies a growth tensor that represents the local addition of mass.
Growth is then modeled through an evolution law for this tensor, typically driven by stress, strain, or other biological or chemical stimuli (e.g., Di Carlo and Quiligotti \cite{DiCarloQuiligotti2002}; Ambrosi and Mollica \cite{Ambrosi2002}; Ambrosi and Guana \cite{Ambrosi2007};  Erlich and Zurlo~\cite{Erlich2024,Erlich2025}).
{Other important} developments incorporated residual stress and geometric incompatibility (e.g., Hoger~\cite{Hoger1986}; Chen and Hoger~\cite{Chen2000}; Epstein and Maugin~\cite{Epstein2000}), revealing how incompatible growth can generate internal stresses even in the absence of external loads.
Such models have also shown that growth may induce mechanical instabilities and pattern formation (e.g., Goriely and Ben Amar~\cite{Goriely2005}; Moulton et al.~\cite{Moulton2013}).  {We also remark that a lagrangian, variatonal theory of growth has been recently proposed in  \cite{Grillo2025} by Grillo, Pastore, and Di Stefano.}

Surface accretion, in contrast to volumetric growth, has been formulated through evolving-domain descriptions, configurational-force balances, and kinetic relations for the interfacial velocity (e.g., Tomassetti et al.~\cite{Tomassetti2016}; Zurlo and Truskinovsky~\cite{Zurlo2017, Zurlo2018, Truskinovsky2019}; Naghibzadeh et al.~\cite{Naghibzadeh2021}; Renzi et al.~\cite{Renzi2024}).
In all these approaches—whether volumetric or interfacial—the growth rate is prescribed through a constitutive evolution law.
This framework has clarified many important consequences of imposed growth fields; however, it does not treat growth as the outcome of a global selection principle.

In the simplest three-dimensional setting, surface growth may be described as follows.
At time $t$, let the body occupy a region
$
R_t \subset \mathbb{R}^3
$ in the reference configuration,
with boundary $\partial R_t$.
If
\[
\mathbf{r} = \mathbf{r}(s^1,s^2,t)
\]
parametrizes $\partial R_t$, the boundary velocity decomposes as
\[
\frac{\partial \mathbf{r}}{\partial t}
= v_n \mathbf{n} + \mathbf{v}_{\mathrm{tan}},
\]
where $\mathbf{n}$ is the outward unit normal, $v_n$ the normal growth velocity, and $\mathbf{v}_{\mathrm{tan}}$ the tangential component.
Only $v_n$ contributes to the geometric evolution; the tangential component merely induces a reparametrization.

Surface growth must satisfy mass conservation at the evolving interface.
Let $\rho$ be the bulk mass density {of the incoming mass} and $J$ the mass flux per unit area supplied to the boundary.
The local interfacial balance then reads
\[
\rho v_n = J.
\]
This relation directly links the normal velocity of the boundary to the incoming mass flux.
Depending on the physical context, one may prescribe the flux $J$, or instead postulate an evolution for $v_n$, with the corresponding flux determined \emph{a posteriori}.

In the latter case, the normal growth velocity may depend on geometric quantities, on the stress state of the body, or on other relevant physical fields.

While prescribing $J$ or specifying a constitutive relation for $v_n$ is appropriate in some physical growth processes, in many biological systems such prescriptions are difficult, if not impossible, to establish from first principles.
One may infer constitutive laws retrospectively from observed growth patterns, but this approach is descriptive rather than predictive.

The idea that living organisms are shaped so as to optimize functional performance is both intuitively plausible and empirically supported.
For example, the morphology of certain organisms enhances feeding efficiency \cite{Liu2025};
the geometry and structure of the rachis in avian feathers achieves an optimized balance between lightness, strength, and flexibility \cite{Sullivan2017};
the spatial distribution of branches and leaves in trees promotes maximal light interception \cite{Johnson2019};
and in both trees and bones, variations in cross-sectional geometry often follow Baud-type profiles that reduce stress concentrations \cite{Mattheck1989}.

Despite these indications of functional optimization in nature, existing continuum theories of growth do not explicitly incorporate an optimality principle.

The main aim of the present manuscript is to show how to formulate a theory of growth driven by an optimality requirement. Our formulation combines ideas from growth mechanics and structural optimization (see Bendsøe and Sigmund~\cite{Bendsoe2003}; Allaire et al.~\cite{Allaire2021}), providing a variational alternative to kinetic growth laws.
The driving mechanism for growth is encoded in an objective functional while equilibrium, mass balance and irreversibility of deposition are imposed as constraints.
Growth thus emerges as a variational selection process rather than from a pointwise evolution law.

Our formulation  bears some resemblance to incremental optimal-design problems for responsive structures (cf. Akerson, Bourdin, and Bhattacharya~\cite{Akerson2022} and Andrini, Noselli, and Lucantonio~\cite{Andrini2022}) and to shape-programming (cf. Ortigosa-Martínez et al.~\cite{OrtigosaMartinez2024}), although the mechanical and variational settings differ.

The emphasis of this work is not on a specific biological or physical application, but rather on the development of the underlying variational framework.  For this reason, we consider a simple  model: a linearly elastic cantilever beam with a rectangular cross-section whose height is allowed to grow. As shown in Section~\ref{sec_sg}, we consider a beam that ``seeks'' to become as rigid as possible; accordingly, growth is driven by the minimization of compliance. Within this variational perspective, we adopt a discrete-time setting, as it proves to be more transparent and conceptually insightful.

Despite its simplicity, the model reveals several structural features of optimality-driven surface growth.
In particular, prestress associated with accretion can alter the convexity properties of the compliance functional.
When the compliance functional is convex, the incremental problem admits a unique minimizer and the evolution is stable.
When convexity is lost due to growth-induced prestrain, nonuniqueness and localization phenomena may arise.
To prevent configurational discontinuities in time, we introduce a quadratic regularization term that penalizes large deviations from the previous configuration. This approach is conceptually related to De Giorgi’s theory of minimizing movements~\cite{DeGiorgi1993, Ambrosio2005} and enables us to derive, from the time-discrete formulation via a formal limiting procedure, a time-continuous limit in the form of a constrained gradient flow.

The paper is organized as follows. In Section~2 we introduce the mechanical model of a layered elastic beam, in which each  layer is characterized by a prescribed prestrain and precurvature. Section~3 presents the variational formulation of superficial growth: we describe the discrete-time incremental scheme, define the compliance functional, and state the associated constrained minimization problems that govern the evolution.

In Section~4 we examine several representative examples and provide numerical results illustrating the qualitative features of the model, with particular emphasis on the role of prestrain and on the convexity properties of the functional. In Section~5  we formally derive a continuous-time formulation as a limit of the discrete incremental scheme. Finally, Section~6 contains a closed-form analytical solution for a simplified case, together with additional technical details.

\section{Mechanics of a layered beam}

\subsection{An elastic beam}\label{sec11}
Consider a cantilever beam of length~$\ell$ with a rectangular cross-section.
The base of the cross-section has unit length into the page, while its height varies along the beam and is described by a function $h_0$.
More precisely, for each $x \in [0,\ell]$, the quantity $h_0(x)$ denotes the height of the cross-section at position~$x$.

We denote by $R_0$ the region occupied by the beam in the reference configuration.
Throughout this work, we restrict ourselves to a two-dimensional setting.
Accordingly, the reference domain is defined as
\[
R_0 := \left\{ (x,y) \in \mathbb{R}^2 \;\middle|\; 0 \le x \le \ell,\; 0 \le y \le h_0(x) \right\}.
\]
\smallskip
\begin{center}
\begin{tikzpicture}[scale=0.5]
\draw[thick] (0,0) -- (0,1.4) -- (10,1.4) -- (10,1);
% Draw the variable thickness line
\draw[thick, smooth] plot[domain=0:10] (\x,{0.01*\x*\x});
%assi
\draw[-latex, blue, thick] (10.5,1.4) -- (13,1.4) node[above] {$x$};
\draw[-latex, blue, thick] (0,-0.3) -- (0,-1.5) node[left] {$y$};
    % Markings and annotations
    % Draw the X arrow
    \draw[-latex] (0,-0.6) -- (7,-0.6) node[midway,above] {$\qquad x$};
    \draw (0,-0.7) -- (0,-0.5);

    \draw[latex-latex] (7,1.4) -- (7,0.5);
   \node[left] at (7, 0.9) {$h_0(x)$};

\begin{scope}[shift={(18,0)}]
\draw (0,0.5) rectangle (0.7,1.4) [thick];
    \draw[latex-latex] (-0.5,1.4) -- (-0.5,0.5);
   \node[left] at (-0.5, 0.9) {$h_0(x)$};
    \draw[latex-latex] (0,-0.5) -- (0.7,-0.5);
   \node[above] at (0.35, -0.5) {$1$};
\end{scope}
\end{tikzpicture}
\end{center}

\smallskip
The beam is subjected to a system of distributed loads {$p(x)\bfe_2$}, which induces at the cross-section $x$ a bending moment $M(x)$. {For equilibrium, as it is well known, we have $M''(x) = p(x)$.}
We assume that the deformation of the beam follows the Bernoulli--Navier hypothesis.
Accordingly, the displacement field $\bfu$ is given by
\begin{equation}\label{u}
\bfu(x,y)
= v(x)\,\bfe_1
+ w(x)\,\bfe_2
- y\,w'(x)\,\bfe_1 ,
\end{equation}
where $v(x)$ and $w(x)$ denote the axial and transverse displacements of the reference line $y=0$, respectively.

\begin{center}
\begin{tikzpicture}[scale=0.5]

% --------------------
% Parameters
% --------------------
\def\L{10}        % beam length
\def\defl{0.02}   % bending intensity

% --------------------
% Fixed support
% --------------------
%\fill[gray!30] (-1,-1) rectangle (0,3);
\draw[thick] (0,0) -- (0,1.4);
%\foreach \y in {-1,-0.7,...,2.8}
%  \draw (-1,\y) -- (0,\y+0.3);
        \foreach \y in {0.0,0.2,...,1.4}
        \draw (0,\y) -- (-0.3,\y+0.1);

% --------------------
% Deformed beam (top edge)
% --------------------
\draw[thick]
  plot[smooth,domain=0:\L]
  (\x,{1.4 - \defl*\x*\x});

% --------------------
% Deformed beam (bottom edge)
% variable thickness h0(x)
% --------------------
\draw[thick]
  plot[smooth,domain=0:(\L-0.15)]
  (\x,{0.01*\x*\x - \defl*\x*\x});

\draw[thick] (\L,1.4 - \defl*\L*\L) --+ (-0.2,-0.4);

% --------------------
% Load (variable, e.g. triangular)
% --------------------
\def\pmax{0.8} % altezza massima grafica del carico

% Frecce con altezza variabile
\foreach \i in {0,1,...,16}{
  \pgfmathsetmacro{\xx}{\L*\i/16}          % x in [0,L]
  \pgfmathsetmacro{\pp}{0.6+\pmax*sin(150*\xx/\L)}    % p(x) = pmax * x/L

  \draw[blue,thick,-latex] (\xx,{2.1+\pp}) -- (\xx,2.1);
}

% Linea di base
\draw[blue] (0,2.1) -- (\L,2.1);

% Inviluppo superiore del carico
\draw[blue,thick]  plot[smooth,domain=0:\L,samples=100]  (\x,{2.1 + 0.6 + \pmax*sin(150*\x/\L)});

\node[right,blue] at (\L,{2.1+\pmax/2}) {$p(x)$};

% --------------------
% Axes
% --------------------
\draw[blue,thick,-latex] (\L+0.5,1.4) -- (\L+3,1.4) node[above] {$x$};
\draw[blue,thick,-latex] (0,-0.5) -- (0,-1.5) node[left] {$y$};

\end{tikzpicture}
\end{center}

In particular, {the unique non-null component $e$ of the strain tensor} is given by
$$
e(x,y)=v'(x)- y\,w''(x)
$$
and setting the axial strain and curvature to be
$$
\varepsilon(x):=v'(x), \quad \kappa(x):=-w''(x),
 $$
we can write
\begin{equation}\label{eps}
e(x,y)= \varepsilon(x)+y\,\kappa(x).
\end{equation}
We assume that the beam is linearly elastic, isotropic, and characterized by Young’s modulus~$E$:
the stress $\sigma$ is given by $\sigma=Ee$. Imposing force and moment balance,
\begin{equation}\label{eqeq}
\int_{0}^{h_0(x)} \sigma(x,y)\, dy = 0,
\qquad
\int_{0}^{h_0(x)} -y \, \sigma(x,y)\, dy = M(x),
\qquad \forall x \in (0,\ell),
\end{equation}
it is found that
\begin{equation}\label{res0}
{\varepsilon(x)= \frac{6 M(x)}{E h_0^2(x)}\quad\mbox{and}\quad \kappa(x)=-\frac{12 M(x)}{E h_0^3(x)}.}
\end{equation}

\subsection{An elastic beam with 1 pre-stressed layer}
Assume that, on the beam considered in the previous section, a layer of additional material is deposited in the reference configuration.
As a consequence, the height of the cross section in the reference configuration at position $x$ becomes $h_1(x)$.
By construction, it holds that
$h_1(x) \ge h_0(x)$ for every $x \in [0,\ell]$.

\begin{center}
\begin{tikzpicture}[scale=0.5]
\draw[thick] (0,-0.5) -- (0,1.4) -- (10,1.4) -- (10,1);

% Original height h_0
\draw[thick, smooth, dashed] plot[domain=0:7] (\x,{0.01*\x*\x});
\draw[thick, smooth] plot[domain=7:10] (\x,{0.01*\x*\x});

% New height h_1
\draw[thick, smooth] plot[domain=0:7] (\x,{-0.5+0.02*\x*\x});

% Axes
\draw[-latex, blue, thick] (10.5,1.4) -- (13,1.4) node[above] {$x$};
\draw[-latex, blue, thick] (0,-0.3) -- (0,-1.5) node[left] {$y$};

% Height annotation
\draw[latex-latex] (3,1.4) -- (3,-0.3);
\node[left] at (3.1,0.7) {$h_1(x)$};
\end{tikzpicture}
\end{center}

Let
\[
R_1 := \left\{ (x,y) \in \mathbb{R}^2 \;\middle|\; 0 \le x \le \ell,\; 0 \le y \le h_1(x) \right\},
\qquad
L_1 := R_1 \setminus R_0 .
\]
Hence, $L_1$ represents the region occupied by the deposited layer in the reference configuration.

We assume that the deposited material is also linearly elastic and isotropic, with Young’s modulus $E$,
but it is not stress free in the reference configuration.
The stress-free configuration is obtained by prescribing a prestrain of the form
\begin{equation}\label{eps1p}
e_1^p(x,y)
= \varepsilon_1^p(x) + y\,\kappa_1^p(x)\qquad \mbox{if }(x,y)\in L_1
\end{equation}
within the region $L_1$.
Here, $\varepsilon_1^p(x)$ and $\kappa_1^p(x)$ denote, respectively, the axial prestrain and the precurvature,
and are assumed to be \emph{known} functions.

With the displacement field given by~\eqref{u} and the corresponding strain defined in~\eqref{eps},
the resulting stress field reads
\[
\sigma(x,y)=
\begin{cases}
E\,e(x,y),
& \text{if } (x,y) \in R_0, \\[4pt]
E\bigl(e(x,y)-e_1^p(x,y)\bigr),
& \text{if } (x,y) \in L_1 .
\end{cases}
\]

As in the previous section, by using~\eqref{eqeq} (with $h_1$ in place of $h_0$) we obtain
\begin{equation}\label{eps1}
\varepsilon=\frac{\varepsilon_1^p\,{h_{1}}-3\,\varepsilon_1^p\,h_{0}-2\,\kappa_1^p\,{h_{0}}^2}{h_{1}^2}(h_1-h_0)+\frac{6\,M}{Eh_{1}^2}
\end{equation}
and
\begin{equation}\label{k1}
\kappa=\frac{4\,\kappa_1^p\,h_{0}^2+\kappa_1^p\,h_{0}\,h_{1}+6\,\varepsilon_1^p\,h_{0}+\kappa_1^p\,h_{1}^2}{h_1^3}(h_1-h_0)-\frac{12\,M}{Eh_{1}^3}.
\end{equation}

\begin{remark}
\mbox{}

\begin{itemize}
\item
The case $\varepsilon_1^p = 0$ and $\kappa_1^p = 0$ corresponds to the deposition of a stress-free layer in the reference configuration, followed by loading of the beam. From \eqref{eps1} and \eqref{k1}, we deduce
\[
\varepsilon= \frac{6M}{E h_1^2}
\quad \text{and} \quad
\kappa = -\frac{12M}{E h_1^3},
\]
which coincide with the expressions for a beam of height $h_1$; compare with \eqref{res0}.

\item
The case
\[
\varepsilon_1^p = \frac{6M}{E h_0^2}
\quad \text{and} \quad
\kappa_1^p = -\frac{12M}{E h_0^3},
\]
 when the prestrain and precurvature are equal to the values obtained in \eqref{res0}, is equivalent to the situation in which the beam is first loaded and a stress-free layer is subsequently deposited on the deformed configuration. Using \eqref{eps1} and \eqref{k1}, one can verify that \eqref{res0} still holds; namely, the newly deposited layer does not contribute to supporting the load in that configuration.
\end{itemize}
\end{remark}

\subsection{An elastic beam with $S$ pre-stressed layers}

The reference configuration is described by a family of functions
$h_i$, for $i = 0,1,\ldots,S$, where $h_i$ denotes the height of the cross
section after the $i$-th deposition. The reference region after the
$i$-th deposition is defined as
\[
R_i := \left\{ (x,y) \in \mathbb{R}^2 \;\middle|\; 0 \le x \le \ell,\;
0 \le y \le h_i(x) \right\}.
\]
The region occupied by the material deposited during the $i$-th
deposition is then given by
\[
L_i := R_i \setminus R_{i-1}.
\]

At each deposition step, the added material is assumed to be linearly
elastic and isotropic, with Young’s modulus $E$, but it is not stress free in the reference configuration. The corresponding stress-free
configurations are obtained by prescribing a prestrain of the form
\begin{equation}\label{epspi}
e_i^p(x,y)
= \varepsilon_i^p(x) + y\,\kappa_i^p(x)
\end{equation}
within the region $L_i$.

Given the displacement field defined in~\eqref{u} and the associated
strain field introduced in~\eqref{eps}, the resulting stress field is
given by
\begin{equation}\label{stressi}
\sigma(x,y)=
\begin{cases}
E\,e(x,y),
& \text{if } (x,y) \in R_0, \\[4pt]
E\bigl(e(x,y)-e_i^p(x,y)\bigr),
& \text{if } (x,y) \in L_i.
\end{cases}
\end{equation}
Again, by using~\eqref{eqeq} we may deduce $\varepsilon$ and $\kappa$.

\section{Surface growth}\label{sec_sg}

We consider the beam described in Section~\ref{sec11}, whose cross-sectional height is denoted by $h_0$ and which is in equilibrium under a prescribed load.

Assuming unit mass density, the total mass of the beam is
\[
    m_0 = \int_{0}^{\ell} h_0(x)\, \mathrm{d}x .
\]
Suppose now that an external agent supplies additional material, which is absorbed by the beam through its surface.
As discussed in the introduction, we adopt a time-discrete formulation rather than a continuous one, as it offers a more intuitive framework.
Accordingly, we assume that the growth process occurs over $S$ discrete time steps.

\noindent
{\sf Step 1.}
Let $m_1$ denote the mass of the beam after the first deposition. {We assume $m_1$ is known.} Clearly, $m_1 > m_0$.
As a consequence of the added material, the cross-sectional height may increase.
We denote by $h_1(x)$ the height of the cross-section at position $x \in [0,\ell]$ after the first deposition.
This function is not known \emph{a priori}.

Assuming that the deposited material diffuses instantaneously over the surface of the beam, the function $h_1$ does not depend on the location of deposition.
Accordingly, $h_1$ may be any function satisfying
\begin{enumerate}
    \item $\displaystyle \int_{0}^{\ell} h_1(x)\, \mathrm{d}x = m_1,$
    \item $h_1(x) \ge h_0(x)$ for every $x \in [0,\ell]$.
\end{enumerate}
The first condition ensures consistency between the total mass and the cross-sectional height of the beam, while the second guarantees that the additional mass is accumulated without any ablation.

{
\begin{remark}
The theory could also contemplate that the deposited material has different mass densities of the original material. If $\rho_0$ and $\rho_1$ are the mass densities of the original and deposited material, we would have
$$
\int_{0}^{\ell} \rho_i h_i(x)\, \mathrm{d}x = m_i \qquad\mbox{ for $i=0,1.$}
$$
\end{remark}
}

We assume that the deposited material, \emph{i.e.}, the material occupying the region where $h_1 > h_0$, has the same Young's modulus as the original material but is not stress-free.
A stress-free configuration is achieved by prescribing a prestrain $e_1^p$ of the form specified in~\eqref{eps1p}.

Imposing equilibrium, \emph{i.e.}, equations~\eqref{eqeq} with $h_1$ in place of $h_0$, the equilibrium strain $e$ has the form~\eqref{eps}, with $\varepsilon$ and $\kappa$ given by~\eqref{eps1} and~\eqref{k1}, respectively.

We now postulate that the growth of the beam, \emph{i.e.}, the choice of the function $h_1$, is governed by the following optimality criterion: \emph{the beam selects $h_1$ so as to maximize its rigidity, or equivalently, to minimize its compliance}.

To make this statement precise, a definition of compliance is required.
In the presence of growth, there is no unique definition; for this reason, we employ a definition of mean compliance commonly used in thermoelasticity, adapted to the present context (see~\cite{Zhang2014, Neiferd2018}). This choice is merely illustrative, as the proposed framework extends to other objective functionals.

The mean compliance associated with the function $h_1$ is defined as
\begin{equation}\label{C}
    C_1(h_1) := \int_{0}^{\ell} \int_{0}^{h_1(x)} E\, e^2(x,y)\, dy dx.
\end{equation}
This  is a non-negative energy-like measure. {The mean compliance is computed based on the total strain $e$, therefore minimizing it simulates the tendency of the structure to  remain as close as possible to the reference configuration.}

Taking into account \eqref{eps} we have
\begin{equation}\label{C0}
C_1(h_1) = E \int_{0}^{\ell} \varepsilon^2(x)h_1(x)+\varepsilon(x)\kappa(x)h_1^2(x)+\frac 1 3 \kappa^2(x)h_1^3(x)\,dx
\end{equation}
with $\varepsilon$ and $\kappa$ given {in terms of  $h_1$} by~\eqref{eps1} and~\eqref{k1}, respectively.
We note that also
$\varepsilon$ and $\kappa$ depend on $h_1$, even if it is not explicitly written.

The problem that determines $h_1$ is
\begin{equation}\label{prob_disc1}
\left\{\begin{aligned}
\min\limits_{h_1} & \ C_1(h_1), \\
&\int_0^\ell h_1(x) \, dx = m_1,\\
&h_1(x)\ge h_{0}(x)\quad x \in [0,\ell].
\end{aligned}
\right.
\end{equation}

In problem \eqref{prob_disc1}, the equilibrium equations are not stated explicitly, but are implicitly imposed and used to derive the expression of the mean compliance functional.

\begin{remark}
To solve the problem numerically, we partition the interval $(0,\ell)$ into $N$ equal subintervals and consider piecewise constant functions on each subinterval.
Specifically, define
\[
\delta := \frac{\ell}{N},
\qquad
x_j := (j-1)\delta,
\qquad j = 1,2,\ldots,N+1.
\]
We consider functions $h_1$ that are constant on each subinterval, namely
\[
h_1(x) = h_1^{(j)}
\quad \forall x \in (x_j, x_{j+1}),
\quad j = 1,2,\ldots,N,
\]
where the values $h_1^{(j)} \in \mathbb{R}$ are unknown constants to be determined.

Let $\varepsilon$ and $\kappa$ be defined as in~\eqref{eps1} and~\eqref{k1}, respectively. For each $j = 1,\ldots,N$, define
\[
c_j := E \Big(
\varepsilon^2(x^c_j) h_1^{(i)}
+ \varepsilon(x^c_j)\kappa(x^c_j) \big(h_1^{(j)}\big){^2}
+ \frac{1}{3}\kappa^2(x^c_j) \big(h_1^{(j)}\big)^3
\Big),
\]
where
\[
x^c_j := \frac{x_j + x_{j+1}}{2}.
\]

The discrete approximation of problem~\eqref{prob_disc1} is given by
\begin{equation}\label{prob_disc1bis}
\left\{
\begin{aligned}
\min_{\{h_1^{(j)}\}} \quad & \delta \sum_{j=1}^N c_j, \\
 \quad
& \delta \sum_{j=1}^N h_1^{(j)} = m_1, \\
& h_1^{(j)} \ge h_0(x^c_j), \quad j = 1,2,\ldots,N.
\end{aligned}
\right.
\end{equation}

This yields a finite-dimensional optimization problem with $N$ unknowns, subject to one scalar equality constraint and $N$ scalar inequality constraints. In MATLAB, problems of this type can be solved using the \texttt{fmincon} command.

\end{remark}

\noindent
{\sf Step $i$.}
Let $m_i$ and $h_i(x)$ denote, respectively, the beam mass and the
cross-sectional height at position $x \in [0,\ell]$ after the $i$-th
deposition.

As before, we assume that the deposited material, \emph{i.e.}, the material
occupying the region where $h_i > h_{i-1}$, has the same Young’s modulus as
the original material but is not stress-free. A stress-free configuration
is obtained by prescribing a prestrain $e_i^p$ of the form given
in~\eqref{epspi}, and consequently, the stress is determined by~\eqref{stressi}.
Imposing equilibrium, \emph{i.e.}, equations~\eqref{eqeq}, the equilibrium
strain $e$, of the form specified in~\eqref{eps}, can be determined.

The mean compliance associated with the function $h_i$ is then defined as
\begin{equation}\label{C}
    C_i(h_i) := \int_{0}^{\ell} \int_{0}^{h_i(x)} E\, e^2(x,y)\, dy\, dx,
\end{equation}
and the problem determining $h_i$  reads
\begin{equation}\label{prob_disc1_3}
\left\{
\begin{aligned}
\min_{h_i} \quad & C_i(h_i), \\
\quad
& \int_0^\ell h_i(x)\, dx = m_i, \\
& h_i(x) \ge h_{i-1}(x), \quad x \in [0,\ell].
\end{aligned}
\right.
\end{equation}

\section{Examples and numerical considerations}

Throughout this section, we consider a beam of length $\ell = 20\,\mathrm{dm}$ with a constant cross-section of height
$h_0 = 0.3\,\mathrm{dm}$, depth $1\, \mathrm{dm}$ into the page, and Young’s modulus $E = 10^5\,\mathrm{N/dm}^2$.

\subsection{Baseline case: $\varepsilon_i^p = 0$ and $\kappa_i^p = 0$ for all $i$}\label{section_00}

We consider the case in which there is no prestrain and precurvature, and assume that the beam is subjected to a uniform load of intensity $p = 0.02\,\mathrm{N/dm}$. This load induces the bending moment
\begin{equation}\label{M0}
    M(x) = \frac{p}{2}(\ell - x)^2.
\end{equation}

\begin{figure}[!hbt]
\centering

\begin{minipage}{0.32\textwidth}
\centering
\begin{overpic}[width=\textwidth]{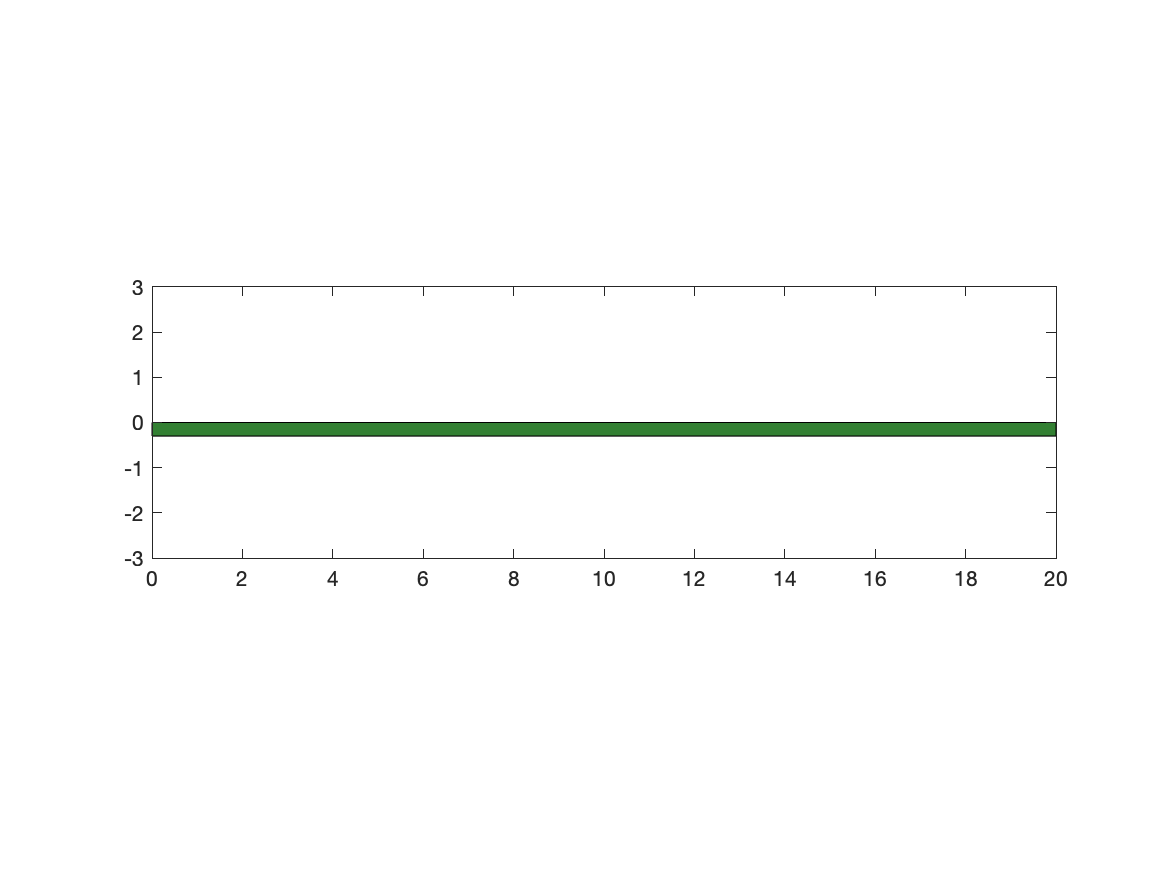}
\put(50,15){\makebox(0,0){$i=0$}}
\end{overpic}
\end{minipage}
\hfill
\begin{minipage}{0.32\textwidth}
\centering
\begin{overpic}[width=\textwidth]{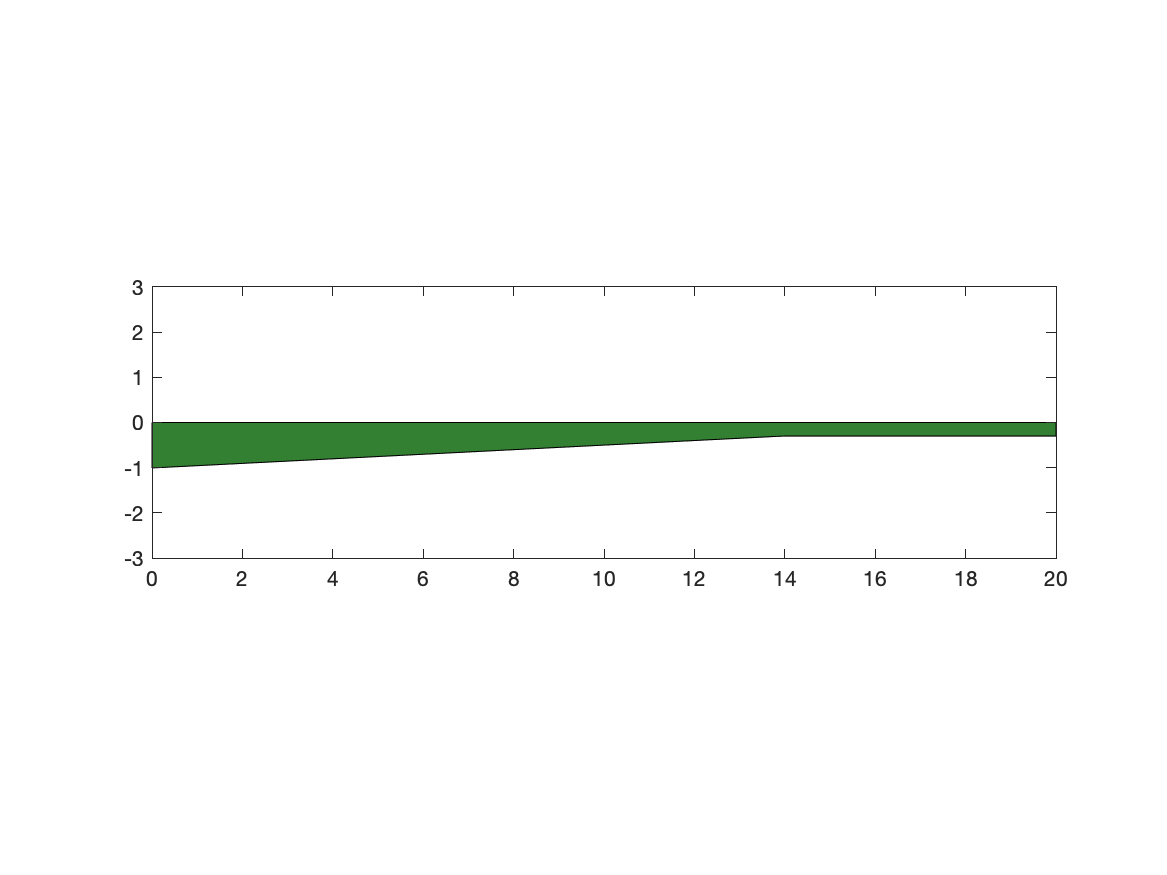}
\put(50,15){\makebox(0,0){$i=5$}}
\end{overpic}
\end{minipage}
\hfill
\begin{minipage}{0.32\textwidth}
\centering
\begin{overpic}[width=\textwidth]{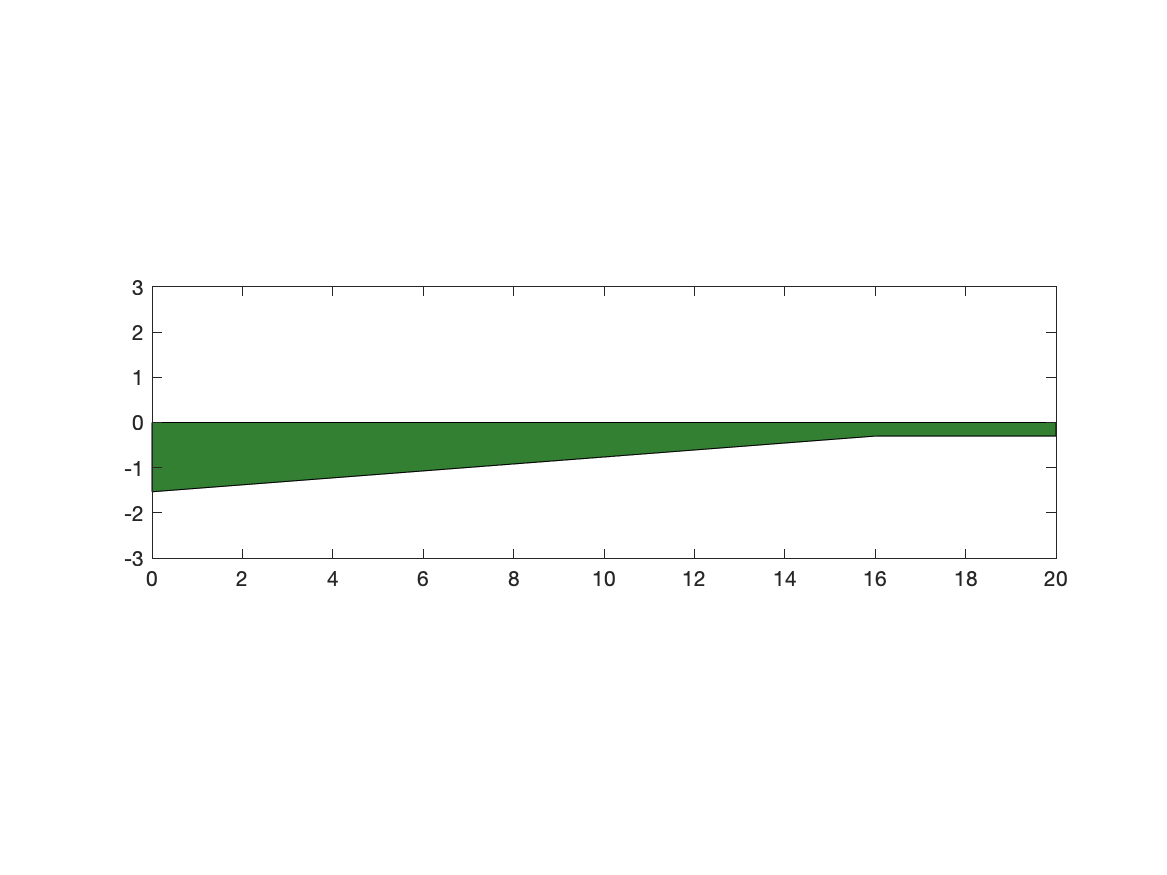}
\put(50,15){\makebox(0,0){$i=10$}}
\end{overpic}
\end{minipage}

\vspace{-0.5cm}
\caption{Beam with $\varepsilon_i^p = 0$ and $\kappa_i^p = 0$ at time steps $i = 0$, $i = 5$, and $i = 10$, with mass
$m_i = m_0 + i\,\Delta m$, where $\Delta m$ is fixed.}
\label{beam1}
\end{figure}

The mean compliance functional is given by \eqref{eps1p}, \eqref{eps1}, \eqref{k1}, \eqref{C}
\begin{equation}\label{Ccase1}
    C_i(h_i) = \int_{0}^{\ell} \frac{12\, M(x)^2}{E\,h_i(x)^3} \, dx,
    \qquad \forall i.
\end{equation}
An analytical solution for $h_i$ can be found (see  Section~\ref{app1}). In particular, it can be shown that the optimal distribution is piecewise-defined: $h_i$ is affine on a subset of $(0,\ell)$ and coincides with $h_{i-1}$ on the complementary subset. This behavior is illustrated in Figure~\ref{beam1}, which reports the numerically obtained results.

\subsection{Case $\varepsilon_i^p = \varepsilon^p$ and $\kappa_i^p = 0$ for all $i$}

We now consider the case in which the precurvature is null and the prestrain is constant. The beam is assumed to be subjected to a bending moment $M(x)$. The mean compliance reads
\begin{equation}\label{C1}
	\begin{aligned}
		C_i(h_i)
		= \int_{0}^{\ell}
		&\frac{3\big(E\varepsilon^p h_0^2 + 2M\big)^2}{E h^3}
		- E(\varepsilon^p)^2 (2h_0 - h_i) \\
		&- \frac{6\varepsilon^p h_0 \big(E\varepsilon^p h_0^2 + 2M\big)}{h_i^2}
		+ \frac{4E(\varepsilon^p)^2 h_0^2}{h_i}
		 \, dx
		\qquad \forall i.
	\end{aligned}
\end{equation}

\subsubsection{Constant bending moment} \label{subsection_cbm}
\begin{figure}[!hbt]
\centering

\begin{minipage}{0.32\textwidth}
\centering
\begin{overpic}[width=\textwidth]{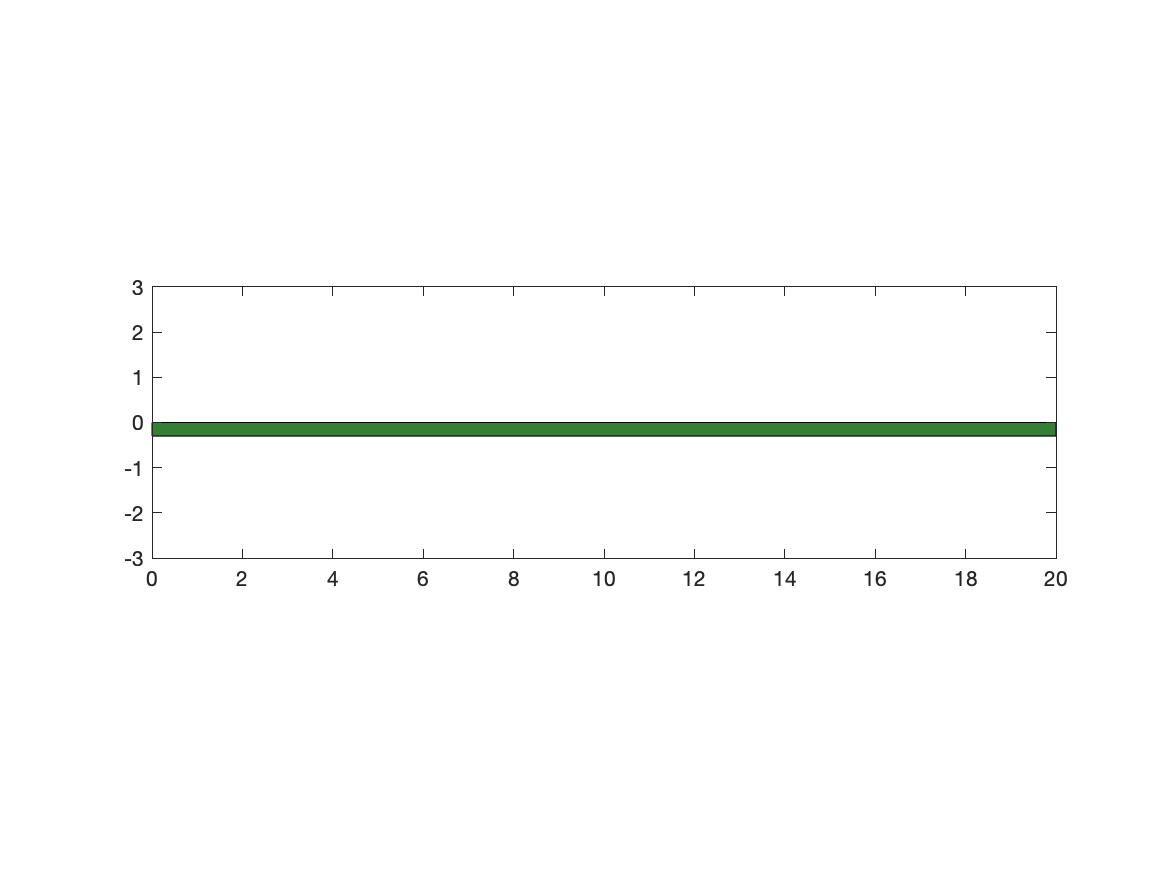}
\put(50,15){\makebox(0,0){$i=0$}}
\end{overpic}
\end{minipage}
\hfill
\begin{minipage}{0.32\textwidth}
\centering
\begin{overpic}[width=\textwidth]{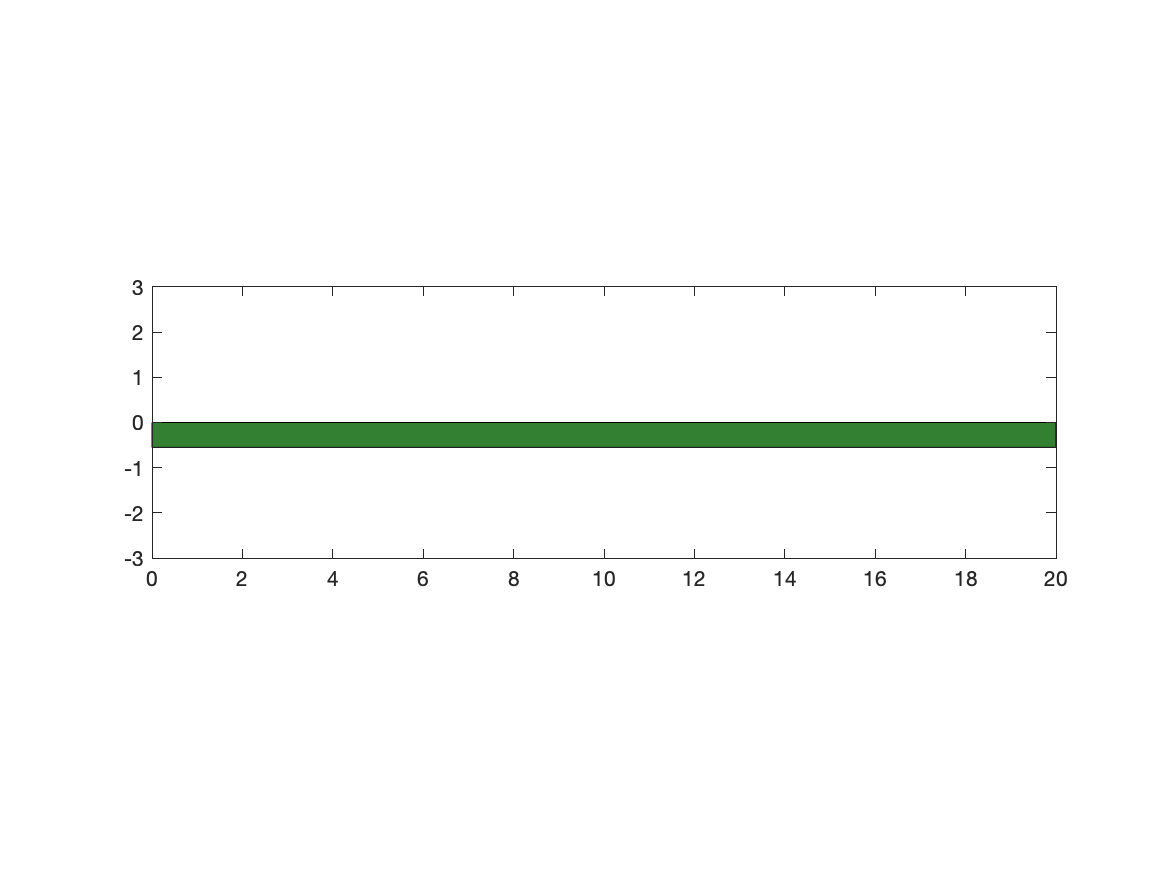}
\put(50,15){\makebox(0,0){$i=5$}}
\end{overpic}
\end{minipage}
\hfill
\begin{minipage}{0.32\textwidth}
\centering
\begin{overpic}[width=\textwidth]{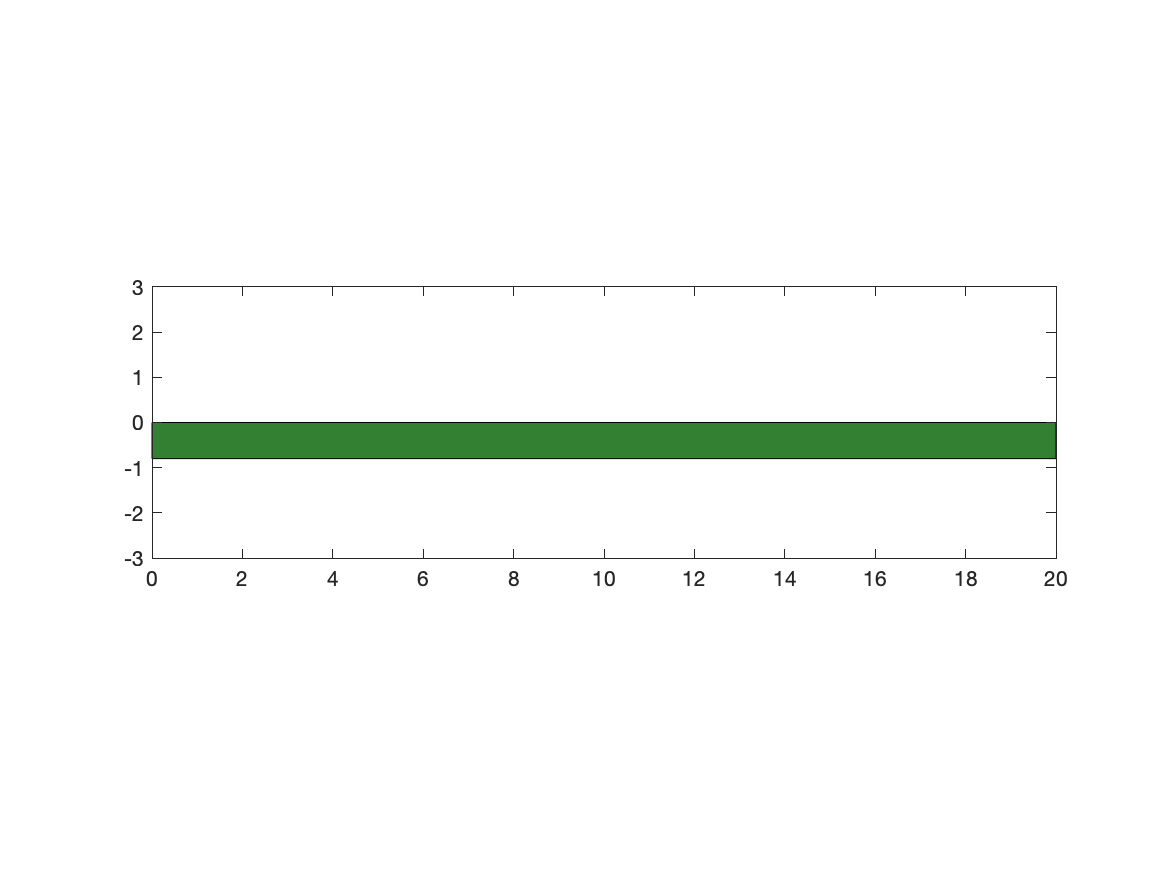}
\put(50,15){\makebox(0,0){$i=10$}}
\end{overpic}
\end{minipage}

\vspace{-0.5cm}
\caption{Constant bending moment: growth of a beam with $\varepsilon^p_i = \varepsilon^p = 0.01$, $\kappa^p_i = 0$, at time steps $i = 0$, $i = 5$, and $i = 10$, with mass
$m_i = m_0 + i\,\Delta m$, where $\Delta m$ is fixed.}
\label{beam2}
\end{figure}

The beam is subjected to a constant bending moment
\( M = 20\,\mathrm{N\,dm} \).
Due to the constancy of the bending moment, uniform growth of the beam is expected. This is confirmed by the results shown in Figure~\ref{beam2}, which reports the numerical results obtained for $\varepsilon^p = 0.01$.

As the beam grows, the height of the cross-section increases, which reduces the deformation induced by the applied bending moment. However, for $\varepsilon^p < 0$, the pre-strain in the added layer tends to bend the existing beam in the same direction as the applied bending moment; consequently, the compliance may not decrease. These considerations suggest that for $\varepsilon^p < 0$, growth may be ineffective in reducing compliance and that the problem may not be well-posed.

\begin{figure}[!hbt]
\centering

\begin{minipage}{0.42\textwidth}
\centering
\begin{overpic}[width=\textwidth]{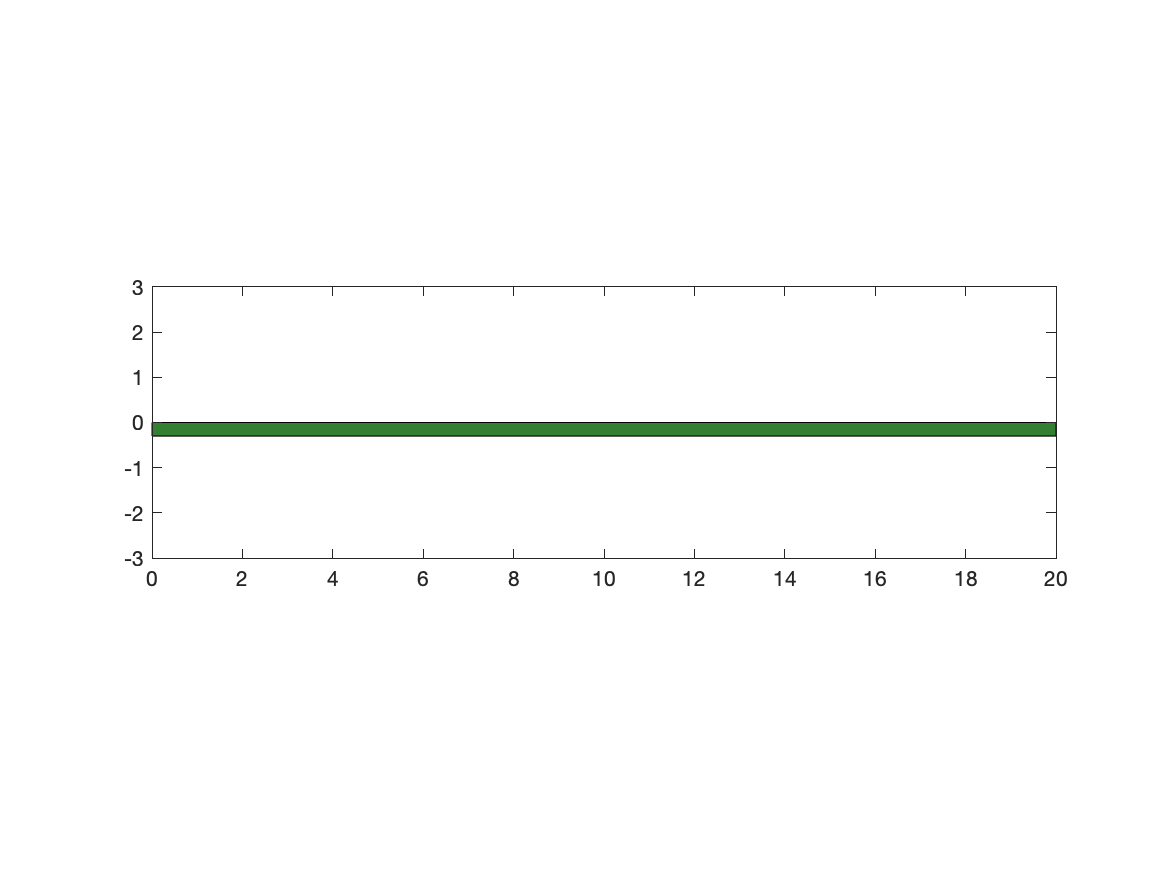}
\put(50,15){\makebox(0,0){$i=0$}}
\end{overpic}
\end{minipage}
\hfill
\begin{minipage}{0.42\textwidth}
\centering
\begin{overpic}[width=\textwidth]{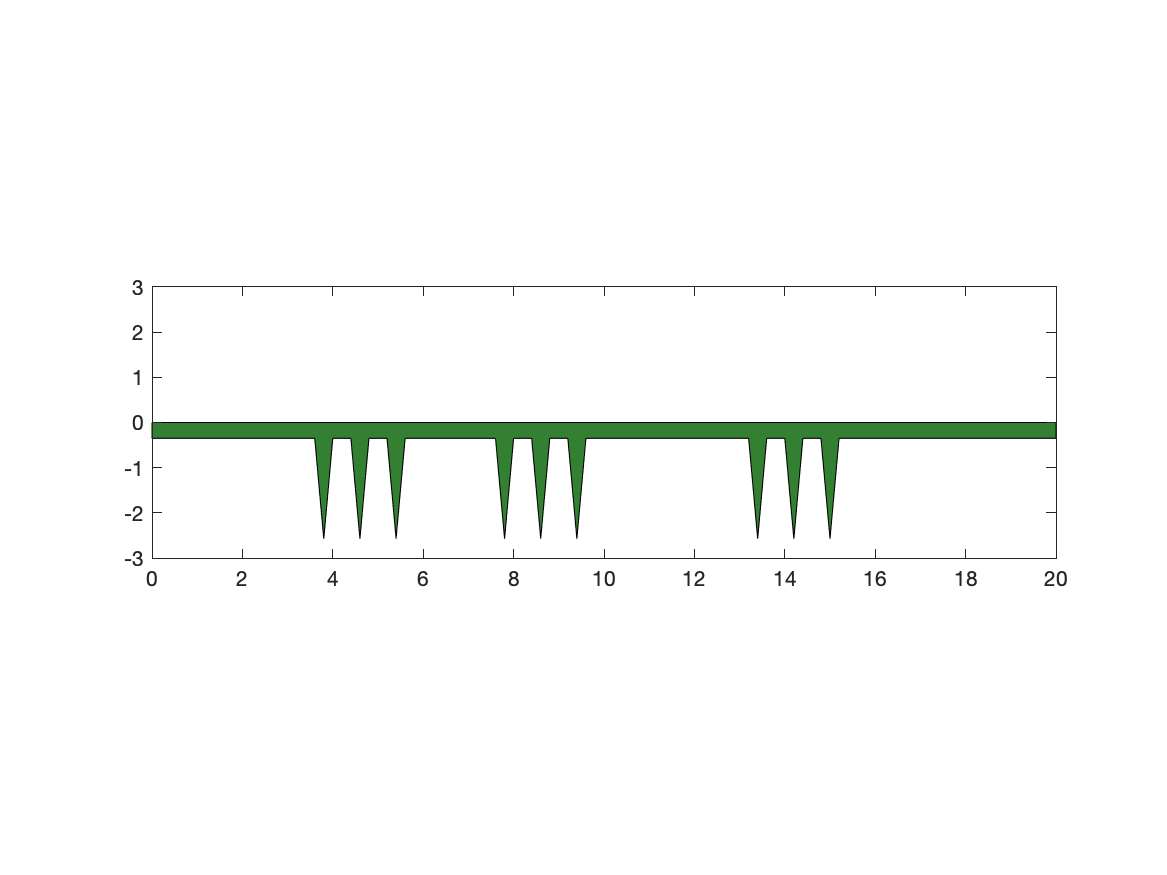}
\put(50,15){\makebox(0,0){$i=5$}}
\end{overpic}
\end{minipage}

\vspace{-0.5cm}
\caption{Constant bending moment: growth of a beam with $\varepsilon^p_i = \varepsilon^p = -0.01$, $\kappa^p_i = 0$, at time steps $i = 0$ and $i = 5$, with mass
$m_i = m_0 + i\,\Delta m$, where $\Delta m$ is fixed.}
\label{beam3}
\end{figure}

The results obtained for $\varepsilon^p = -0.01$, shown in Figure~\ref{beam3}, indicate that growth becomes concentrated at spurious locations. This behavior is symptomatic of numerical issues, possibly related to the lack of convexity of the integrand.

To further analyze this behavior, we introduce the following dimensionless\footnote{We omit writing the unitary depth into the page.} quantities:
\[
\hbar(x) = \frac{h_i(x)}{h_0(x)},
\qquad
\eta(x) = \frac{M(x)}{E h_0^{2}(x)\varepsilon^p(x)} ,
\]
which allow us to rewrite the mean compliance~\eqref{C1} as
\[
C_i(h_i)
= \int_{0}^{\ell} E \bigl(\varepsilon^p(x)\bigr)^2 h_0(x)\,
f(\hbar(x)) \, dx,
\qquad \forall i,
\]
where
\[
f(\hbar)
:= \frac{12\eta^2 - 12\eta \hbar + 12\eta + \hbar^4 - 2\hbar^3 + 4\hbar^2 - 6\hbar + 3}{\hbar^3}.
\]

The integrand of the mean compliance is convex if and only if the function $f$ is convex. Its second derivative is given by
\[
f''(\hbar)
= \frac{4\left(6\eta - \hbar + 3\right)\left(6\eta - 2\hbar + 3\right)}{\hbar^5},
\]
and therefore $f'' \le 0$ for $\hbar_m \le \hbar \le \hbar_M$, where
\[
\hbar_m := \min\!\left(6\eta + 3,\; \tfrac12(6\eta + 3)\right),
\qquad
\hbar_M := \max\!\left(6\eta + 3,\; \tfrac12(6\eta + 3)\right).
\]

Taking into account that $\hbar \ge 1$ and using the fixed geometrical and mechanical parameters, we obtain
\[
\text{for } \varepsilon^p = -0.01,
\qquad
\min f'' = f''(1) = -0.89,
\]
whereas
\[
\text{for } \varepsilon^p = 0.01,
\qquad
\min f'' = f''(2.56) = -0.05.
\]

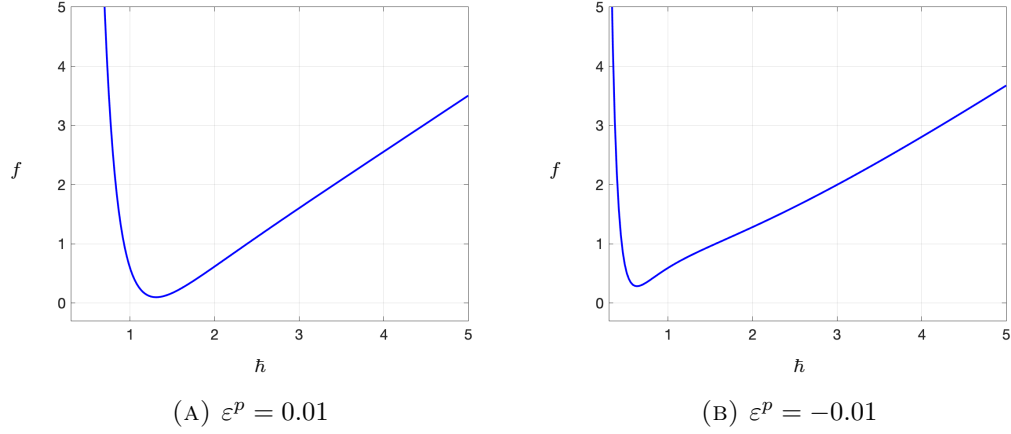
\begin{figure}[!hbt]
\centering

% ---- left panel ----
\begin{subfigure}{0.45\textwidth}
\centering
\begin{tikzpicture}
\node (imgA) {\includegraphics[width=\linewidth]{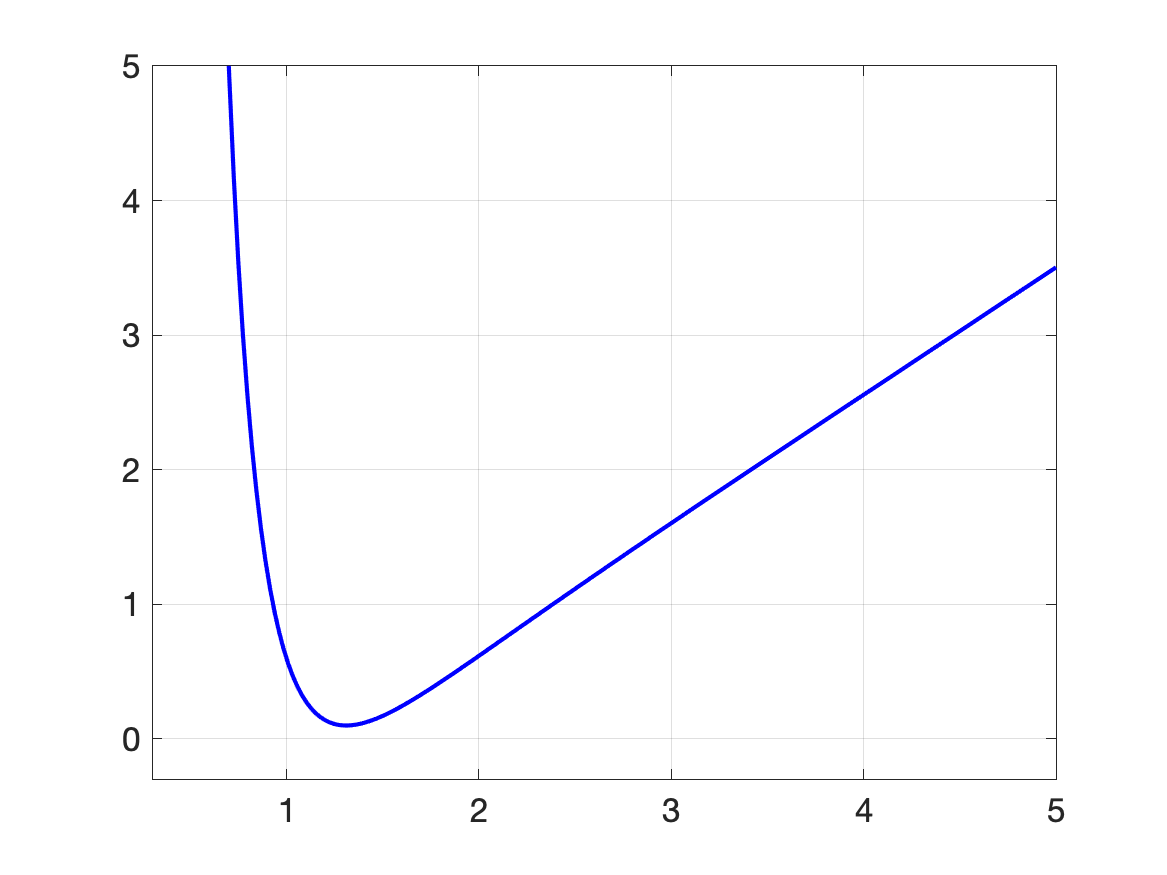}};
% y-label for this panel (change -6mm if needed)
\node at ([xshift=3mm]imgA.west) {\tiny $f$};
% x-label for this panel (change -6mm if needed)
\node at ([yshift=1mm]imgA.south) {\tiny $\hbar$};
\end{tikzpicture}
\caption{$\varepsilon^p = 0.01$}
\end{subfigure}
\hspace{0.1cm}
% ---- right panel ----
\begin{subfigure}{0.45\textwidth}
\centering
\begin{tikzpicture}
\node (imgB) {\includegraphics[width=\linewidth]{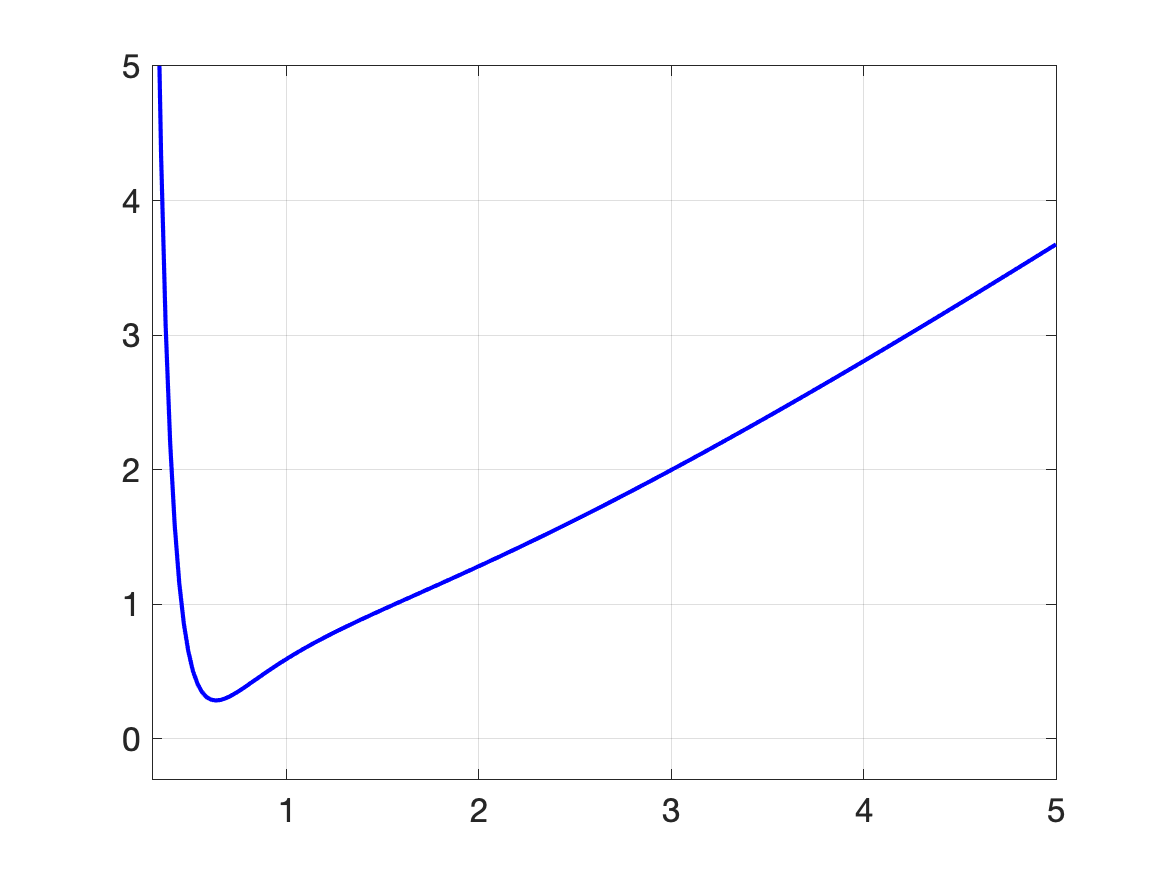}};
\node at ([xshift=3mm]imgB.west) {\tiny $f$};
\node at ([yshift=1mm]imgB.south) {\tiny $\hbar$};
\end{tikzpicture}
\caption{$\varepsilon^p = -0.01$}
\end{subfigure}

\caption{Plot of the function $f$ for  $\varepsilon^p = 0.01$ and $\varepsilon^p = -0.01$.}
\label{f1}
\end{figure}

\begin{figure}[!hbt]
\centering

% ---- left panel ----
\begin{subfigure}{0.45\textwidth}
\centering
\begin{tikzpicture}
\node (imgA) {\includegraphics[width=\linewidth]{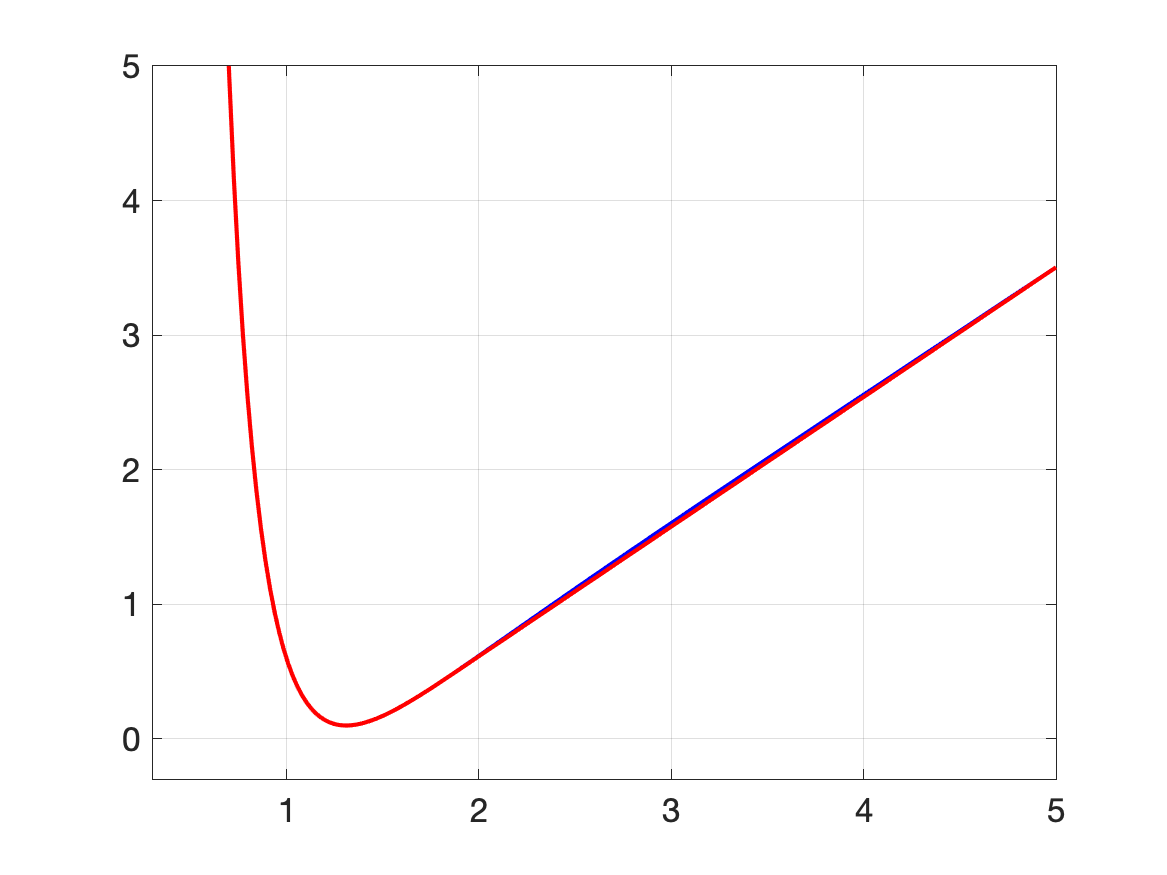}};
% y-label for this panel (change -6mm if needed)
\node at ([xshift=3mm]imgA.west) {\tiny $f^{**}$};
% x-label for this panel (change -6mm if needed)
\node at ([yshift=1mm]imgA.south) {\tiny $\hbar$};
\end{tikzpicture}
\caption{$\varepsilon^p = 0.01$}
\end{subfigure}
\hspace{0.1cm}
% ---- right panel ----
\begin{subfigure}{0.45\textwidth}
\centering
\begin{tikzpicture}
\node (imgB) {\includegraphics[width=\linewidth]{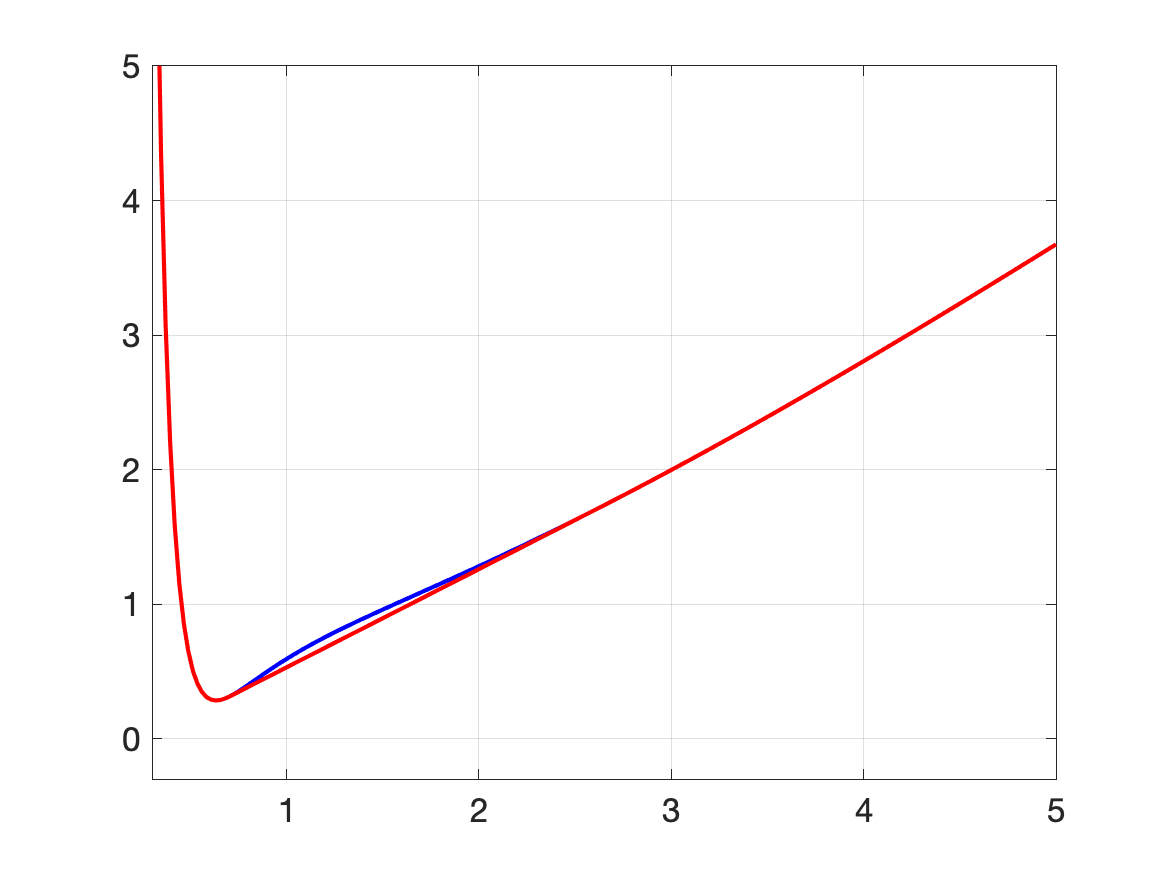}};
\node at ([xshift=3mm]imgB.west) {\tiny $f^{**}$};
\node at ([yshift=1mm]imgB.south) {\tiny $\hbar$};
\end{tikzpicture}
\caption{$\varepsilon^p = -0.01$}
\end{subfigure}

\caption{Plot of the function $f$ (blue) and its convex envelope (red) for $\varepsilon^{p} = 0.01$ and
	$\varepsilon^{p} = -0.01$. For $\varepsilon^{p} = 0.01$ the convex envelope almost coincides with $f$, while for
	$\varepsilon^{p} = -0.01$ the convex envelope and $f$ quite differ for $\hbar$ in $(0.8,2.3)$.}
\label{f2}
\end{figure}
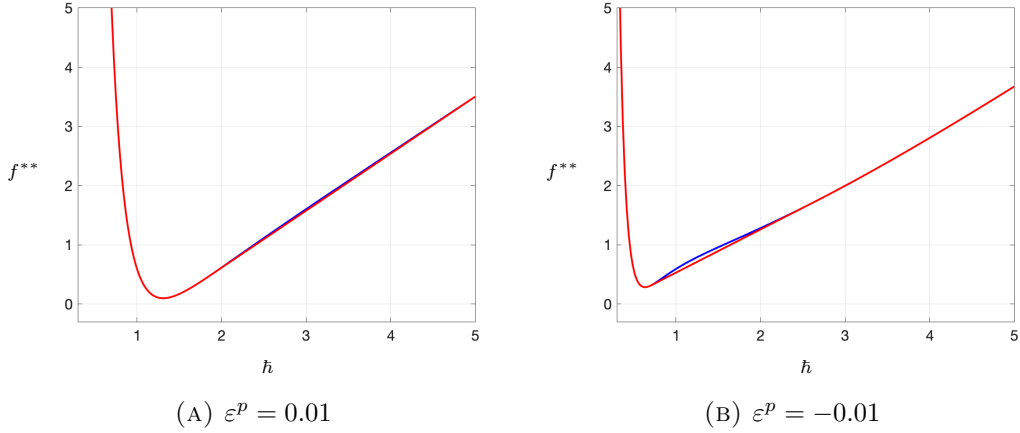

Thus, the integrand of the mean compliance is not convex for $\varepsilon^p = \pm 0.01$. However, in the case $\varepsilon^p = 0.01$ the loss of convexity is ``sufficiently mild'', and more likely numerically $f$ behaves like a convex function, see Figure \ref{f2}. In contrast, for $\varepsilon^p = -0.01$ the stronger lack of convexity leads to a loss of uniqueness of the solution and as a consequence to numerical instabilities.

In general, in problems where uniqueness of the solution is not guaranteed,
the growth process generated by \eqref{prob_disc1} may be highly ``discontinuous'': at a given time step,
the selected solution (among the set of solutions) may be quite
``far'' from the solution chosen at the previous step.
To prevent this issue, we replace problem~\eqref{prob_disc1} with
\begin{equation}\label{prob_disc1_cor}
\left\{
\begin{aligned}
\min_{h_i} \quad
& C_i(h_i) + \frac{1}{2\tau} \int_0^\ell |h_i - h_{i-1}|^2 \, dx, \\
& \int_0^\ell h_i(x)\, dx = m_i, \\
& h_i(x) \ge h_{i-1}(x), \quad x \in [0,\ell],
\end{aligned}
\right.
\end{equation}
where $\tau > 0$ is a regularization parameter. The additional term, weighted by $\frac{1}{2\tau}$, penalizes large deviations of $h_i$ from the previous configuration $h_{i-1}$. Heuristically, if multiple solutions exist, the penalization term selects, among them, the one closest to the solution chosen at the previous step. Moreover, if $\tau$ is sufficiently small, the resulting functional becomes strictly convex and, as a consequence, the solution is unique and does not present spurious localizations. {We remark that the penalization term resembles a time-discrete kinetic energy.}

Problem~\eqref{prob_disc1_cor} has the same formulation as problems related to the notion of minimizing movements introduced by De Giorgi \cite{DeGiorgi1993}. We follow the notation of Braides \cite{Braides2014} for the constant $\frac{1}{2\tau}$.

\begin{figure}[!hbt]
\centering

\begin{minipage}{0.44\textwidth}
\centering
\begin{overpic}[width=\textwidth]{ekedn_0.png}
\put(50,15){\makebox(0,0){$i=0$}}
\end{overpic}
\end{minipage}
\hfill
\begin{minipage}{0.44\textwidth}
\centering
\begin{overpic}[width=\textwidth]{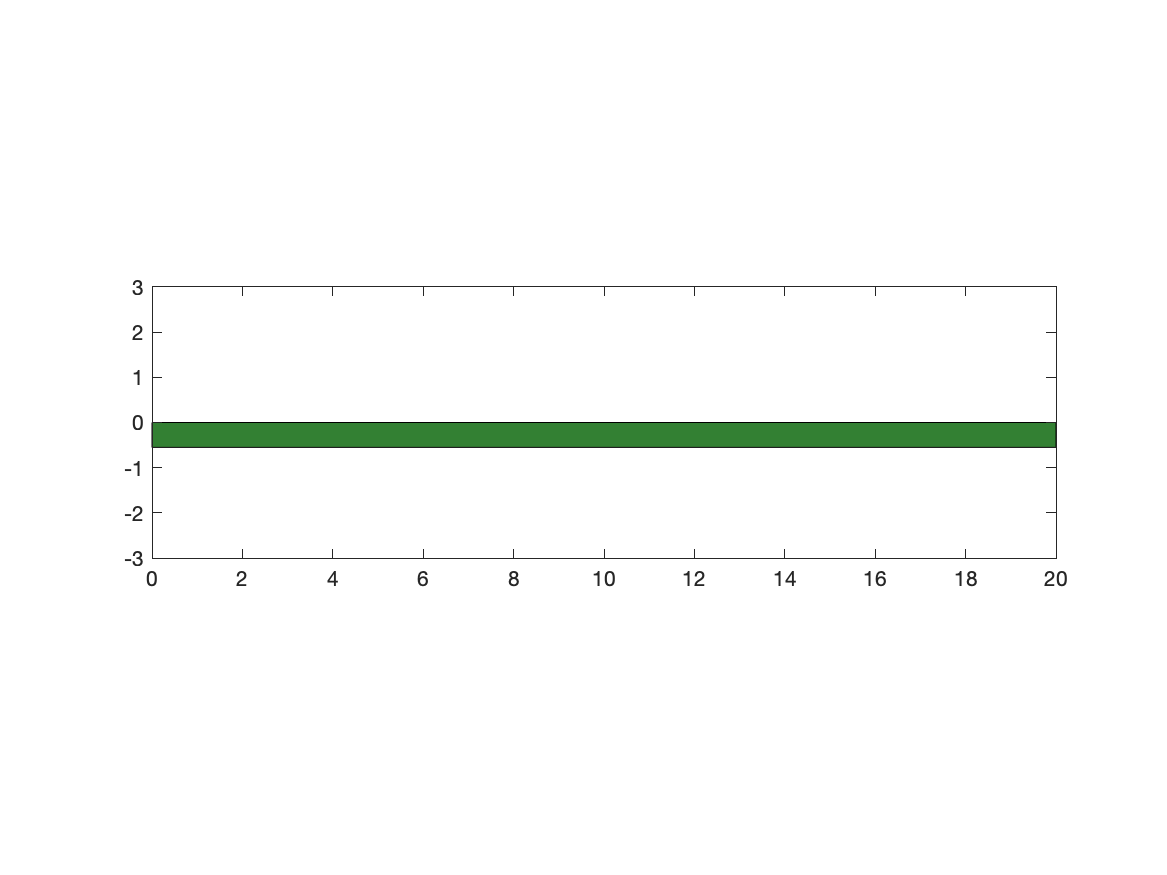}
\put(50,15){\makebox(0,0){$i=5$}}
\end{overpic}
\end{minipage}

\vspace{-0.5cm}
\caption{Constant bending moment: growth of a beam with $\varepsilon^p = -0.01$, $\kappa^p = 0$, and regularization parameter $\tau = 0.01$ at time steps $i = 0$ and $i = 5$, with mass
$m_i = m_0 + i\,\Delta m$, where $\Delta m$ is fixed.}
\label{beam4}
\end{figure}

The results obtained with $\tau = 0.01$ and $\varepsilon^p = -0.01$ are shown in Figure~\ref{beam4}, where no numerical instabilities are observed. Compare Fig. \ref{beam4} with Fig. \ref{beam3}.

\begin{remark}\label{rem3}
If problem~\eqref{prob_disc1_cor} is replaced by
\begin{equation}\label{prob_disc1_cor_less}
\left\{
\begin{aligned}
\min_{h_i} \quad
& C_i(h_i) + \frac{1}{2\tau} \int_0^\ell |h_i - h_{i-1}|^2 \, dx, \\
& \int_0^\ell h_i(x)\, dx \le m_i, \\
& h_i(x) \ge h_{i-1}(x), \quad x \in [0,\ell],
\end{aligned}
\right.
\end{equation}
then $m_i$ no longer represents the mass after step~$i$, but rather the maximum mass that the beam is allowed to have at step~$i$.
In other words, the beam may decide whether or not to absorb the material supplied to it.

If $\varepsilon^p = -0.01$, the beam will not absorb material, since adding new material increases the beam deflection induced by the bending moment and, consequently, the growth process would increase the mean compliance.
\end{remark}

\subsubsection{Parabolic bending moment}

We consider the case in which the beam is subjected to a uniform load of intensity $p$, inducing the bending moment given by \eqref{M0}. For $\varepsilon^p = 0.01$, the cross-sectional height increases, but its  variation along the $x$ direction is `small'. For this reason, instead of plotting the regions $R_i$ as done, for instance, in Figure~\ref{beam4}, we simply display the height function.

\begin{figure}[!hbt]
\centering

% ---- left panel ----
\begin{subfigure}{0.45\textwidth}
\centering
\begin{tikzpicture}
\node (imgA) {\includegraphics[width=\linewidth]{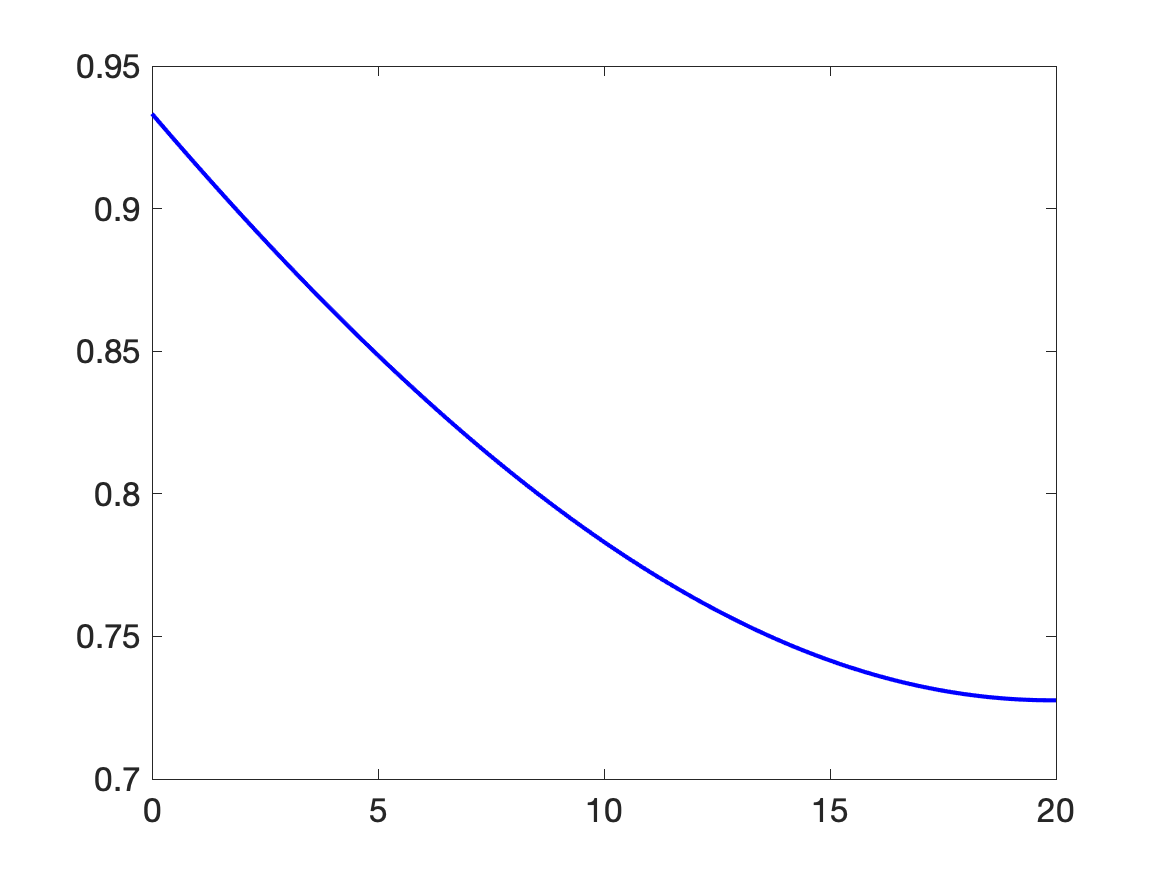}};
% y-label for this panel (change -6mm if needed)
\node at ([xshift=3mm]imgA.west) {\tiny $h$};
% x-label for this panel (change -6mm if needed)
\node at ([yshift=1mm]imgA.south) {\tiny $x$};
\end{tikzpicture}
\caption{$\tau=+\infty$}
\end{subfigure}
\hspace{0.1cm}
% ---- right panel ----
\begin{subfigure}{0.45\textwidth}
\centering
\begin{tikzpicture}
\node (imgB) {\includegraphics[width=\linewidth]{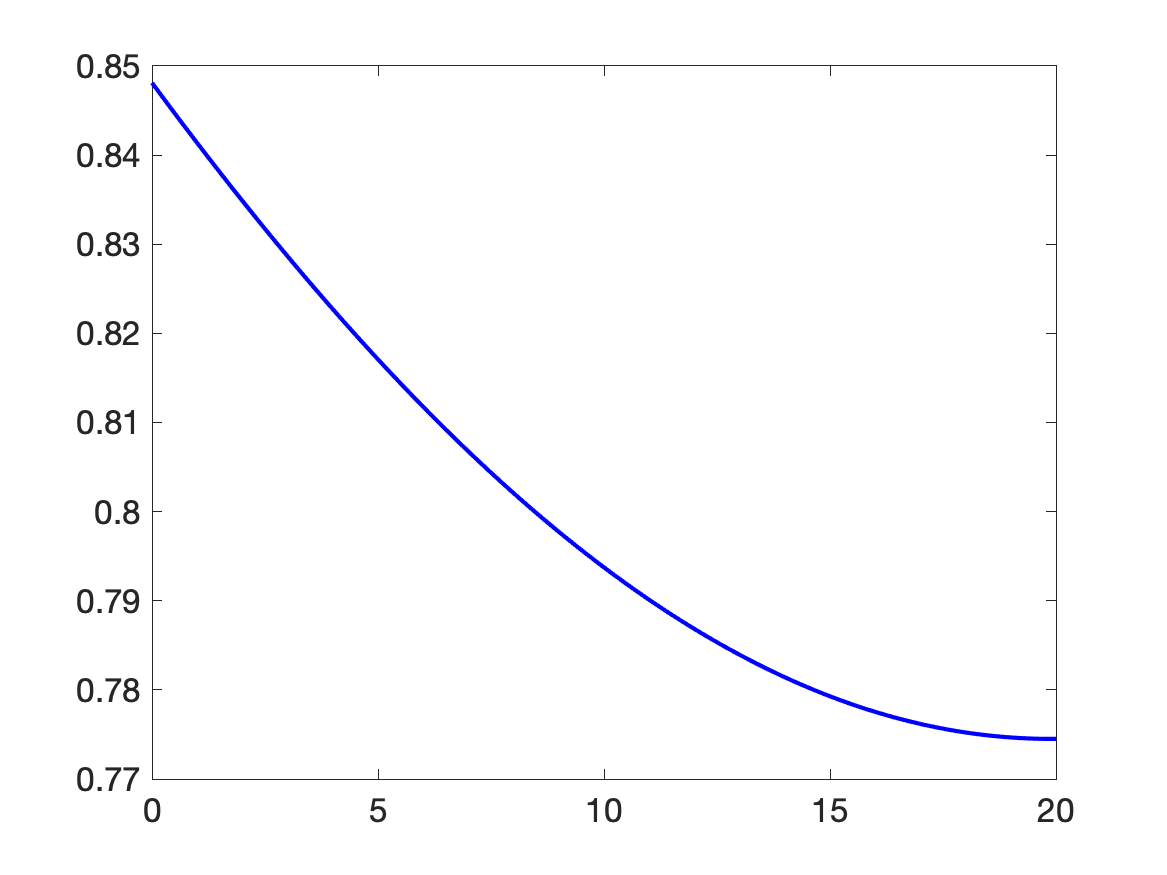}};
\node at ([xshift=3mm]imgB.west) {\tiny $h$};
\node at ([yshift=1mm]imgB.south) {\tiny $x$};
\end{tikzpicture}
\caption{$\tau=0.01$}
\end{subfigure}
\caption{Parabolic bending moment: height of the beam cross-section for $p = 0.02\,\mathrm{N/dm}$, $\varepsilon^p = 0.01$, $\kappa^p=0$,
shown for $\tau = +\infty$ and $\tau = 0.01$ after 10 time steps.}
\label{beamp1}
\end{figure}
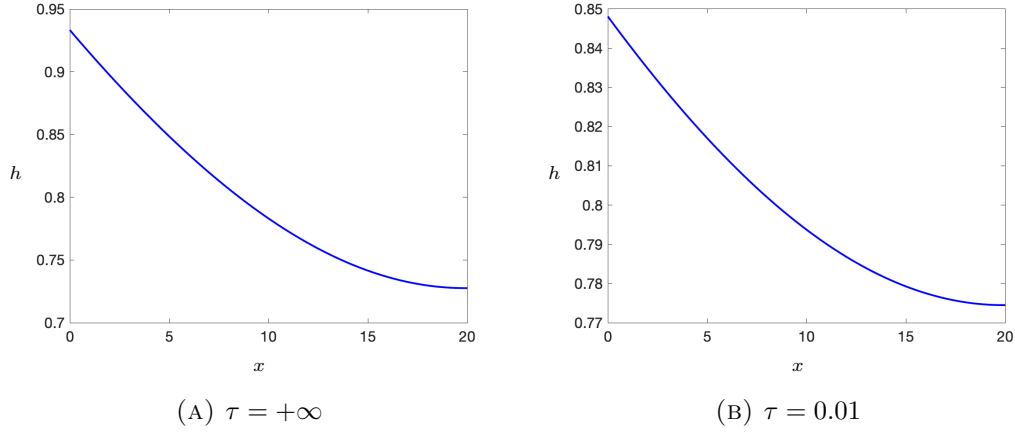
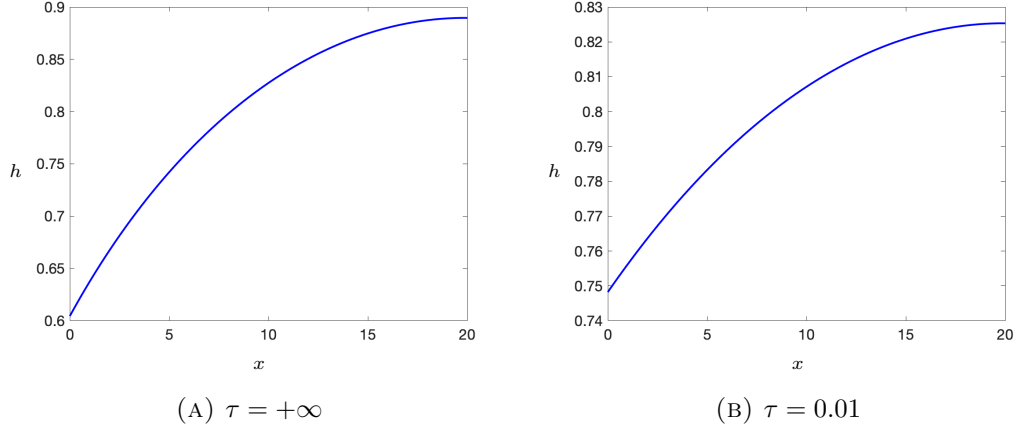

In Figure~\ref{beamp1} we consider the load $p = 0.02\,\mathrm{N/dm}$, as in Section~\ref{section_00}, and $\varepsilon^p = 0.01$. We observe that the solution obtained with $\tau = +\infty$, that is, without the penalization term introduced in \eqref{prob_disc1_cor}, tends to allocate more material towards the fixed end of the cantilever beam compared to the solution obtained with $\tau = 0.01$.

\begin{figure}[!hbt]
\centering

% ---- left panel ----
\begin{subfigure}{0.45\textwidth}
\centering
\begin{tikzpicture}
\node (imgA) {\includegraphics[width=\linewidth]{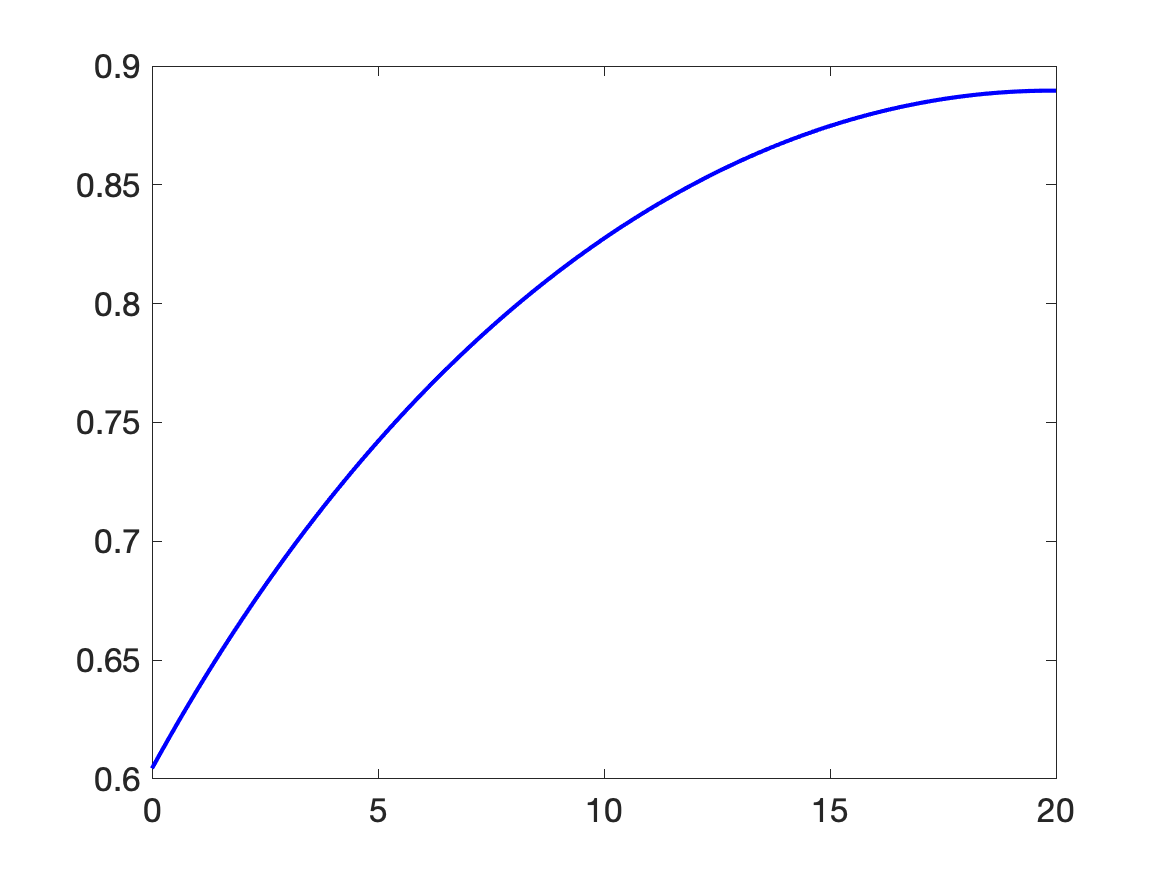}};
% y-label for this panel (change -6mm if needed)
\node at ([xshift=3mm]imgA.west) {\tiny $h$};
% x-label for this panel (change -6mm if needed)
\node at ([yshift=1mm]imgA.south) {\tiny $x$};
\end{tikzpicture}
\caption{$\tau=+\infty$}
\end{subfigure}
\hspace{0.1cm}
% ---- right panel ----
\begin{subfigure}{0.45\textwidth}
\centering
\begin{tikzpicture}
\node (imgB) {\includegraphics[width=\linewidth]{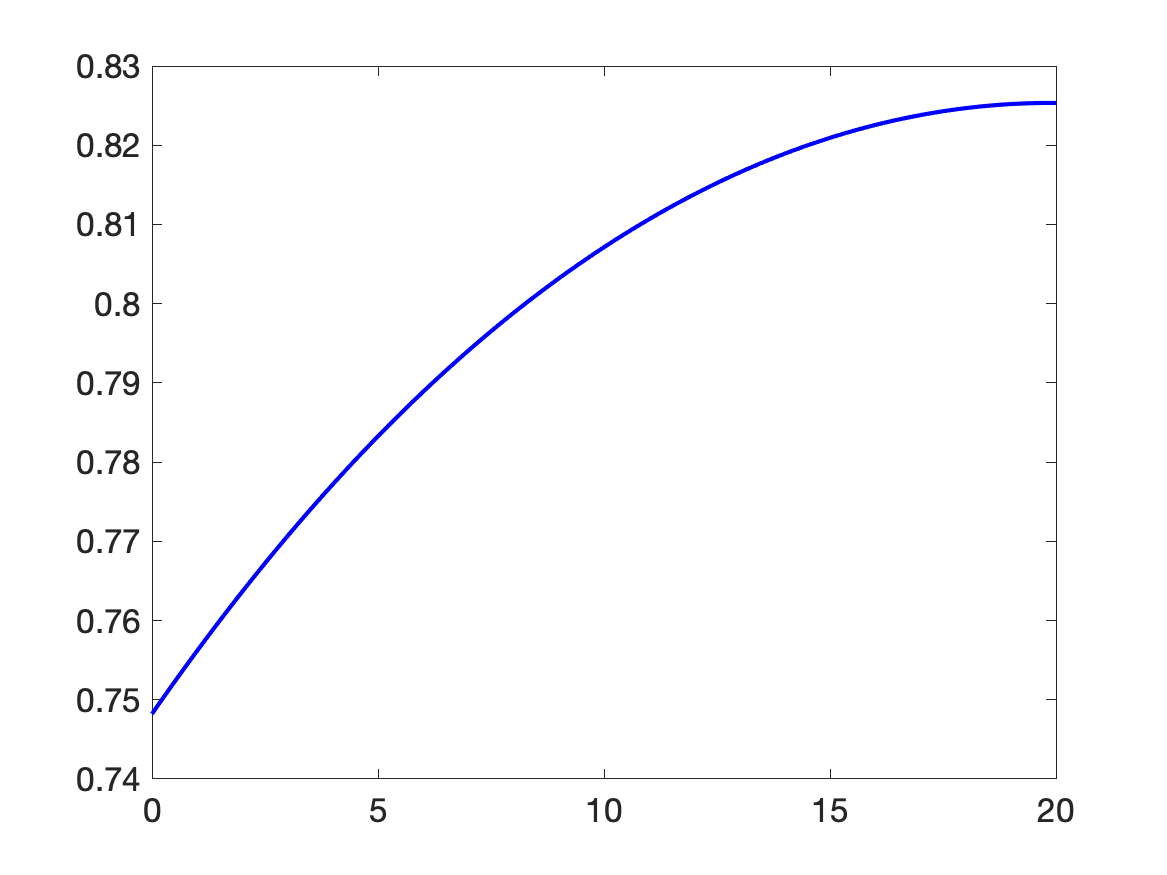}};
\node at ([xshift=3mm]imgB.west) {\tiny $h$};
\node at ([yshift=1mm]imgB.south) {\tiny $x$};
\end{tikzpicture}
\caption{$\tau=0.01$}
\end{subfigure}

\caption{Parabolic bending moment: height of the beam cross-section for $p = 0.02\,\mathrm{N/dm}$, $\varepsilon^p = -0.01$, $\kappa^p=0$,
shown for $\tau = +\infty$ and $\tau = 0.01$ after 10 time steps.}
\label{beamp2}
\end{figure}

Still considering a load $p = 0.02\, \mathrm{N/dm}$ but $\varepsilon^p = -0.01$, a similar situation is observed, except that in this case more material is distributed toward the free end of the beam, see Figure~\ref{beamp2}. We notice that for $\tau=+\infty$ there are no numerical instabilities as those observed in Section
\ref{subsection_cbm}.

\begin{figure}[!hbt]
\centering

% ---- left panel ----
\begin{subfigure}{0.45\textwidth}
\centering
\begin{tikzpicture}
\node (imgA) {\includegraphics[width=\linewidth]{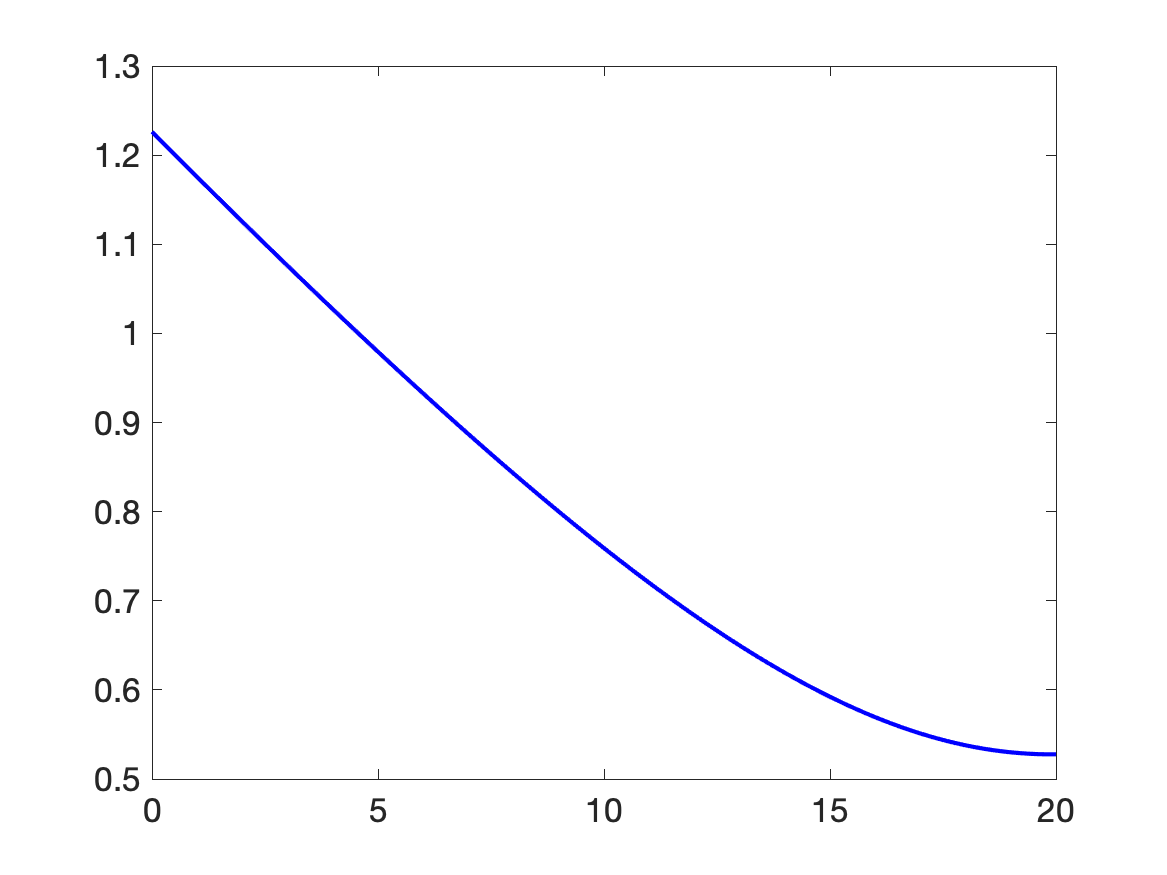}};
% y-label for this panel (change -6mm if needed)
\node at ([xshift=3mm]imgA.west) {\tiny $h$};
% x-label for this panel (change -6mm if needed)
\node at ([yshift=1mm]imgA.south) {\tiny $x$};
\end{tikzpicture}
\caption{$\tau=+\infty$}
\end{subfigure}
\hspace{0.1cm}
% ---- right panel ----
\begin{subfigure}{0.45\textwidth}
\centering
\begin{tikzpicture}
\node (imgB) {\includegraphics[width=\linewidth]{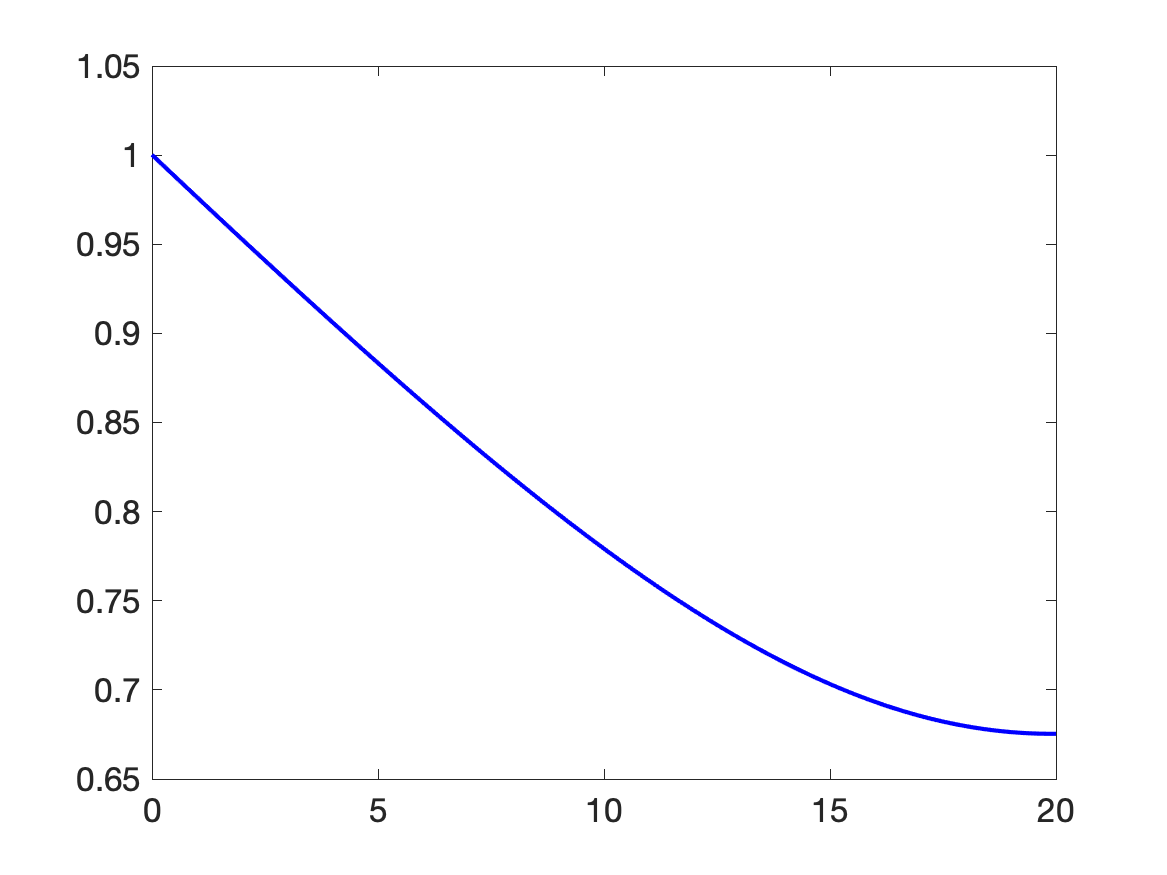}};
\node at ([xshift=3mm]imgB.west) {\tiny $h$};
\node at ([yshift=1mm]imgB.south) {\tiny $x$};
\end{tikzpicture}
\caption{$\tau=0.01$}
\end{subfigure}

\caption{Parabolic bending moment: height of the beam cross-section for $p = 0.1\,\mathrm{N/dm}$, $\varepsilon^p = 0.01$, $\kappa^p=0$,
shown for $\tau = +\infty$ and $\tau = 0.01$ after 10 time steps.}
\label{beamp3}
\end{figure}
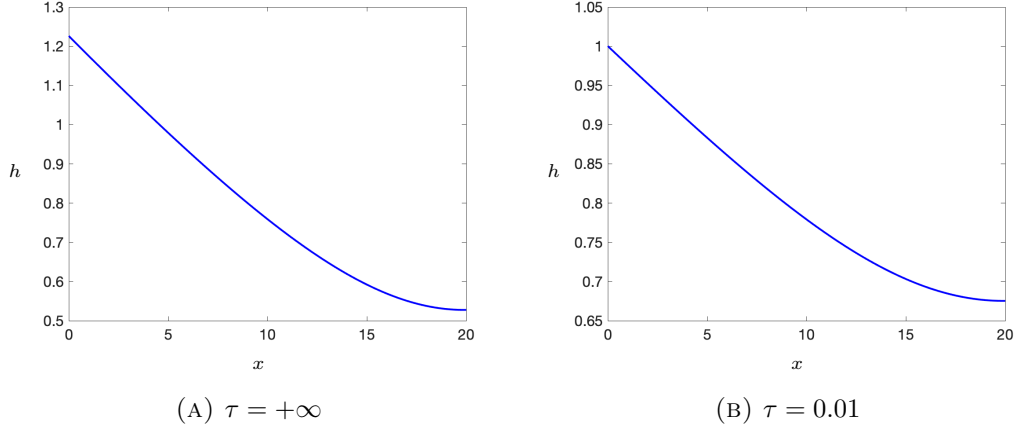

With $p = 0.1\,\mathrm{N/dm}$ the bending moment in the fixed support of the cantilever beam is equal to that considered
in Section \ref{subsection_cbm}. With this value of the distributed load and $\varepsilon^p=0.01$ we find solutions (Figure \ref{beamp3}) similar to those reported in Figure \ref{beamp1}, only that more material is distributed towards the fixed end of the cantilever.

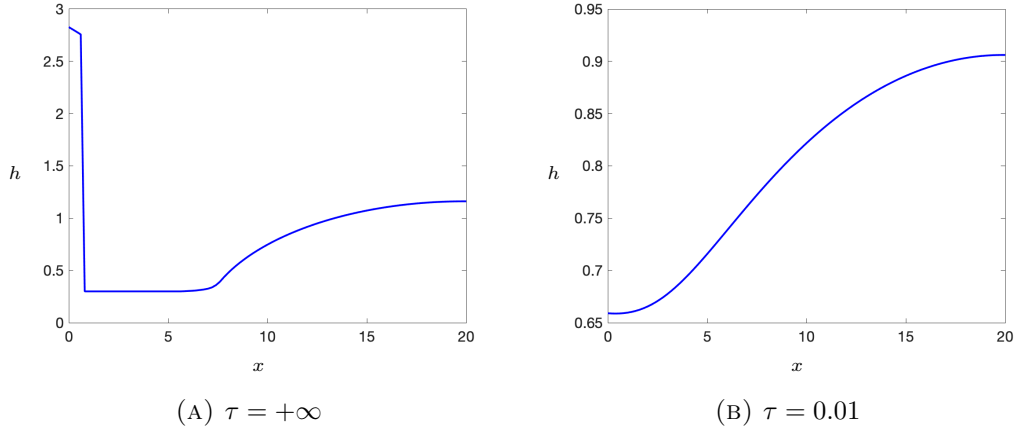
\begin{figure}[!hbt]
\centering

% ---- left panel ----
\begin{subfigure}{0.45\textwidth}
\centering
\begin{tikzpicture}
\node (imgA) {\includegraphics[width=\linewidth]{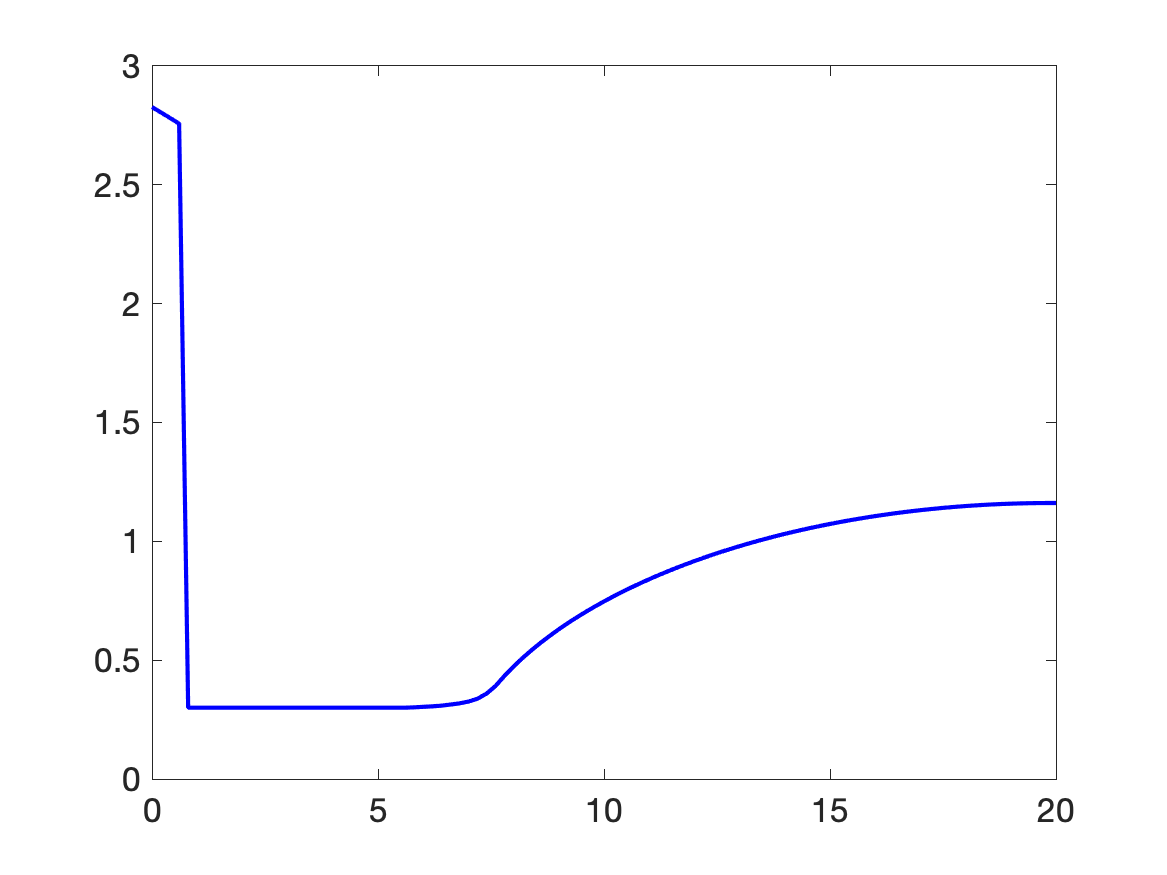}};
% y-label for this panel (change -6mm if needed)
\node at ([xshift=3mm]imgA.west) {\tiny $h$};
% x-label for this panel (change -6mm if needed)
\node at ([yshift=1mm]imgA.south) {\tiny $x$};
\end{tikzpicture}
\caption{$\tau=+\infty$}
\end{subfigure}
\hspace{0.1cm}
% ---- right panel ----
\begin{subfigure}{0.45\textwidth}
\centering
\begin{tikzpicture}
\node (imgB) {\includegraphics[width=\linewidth]{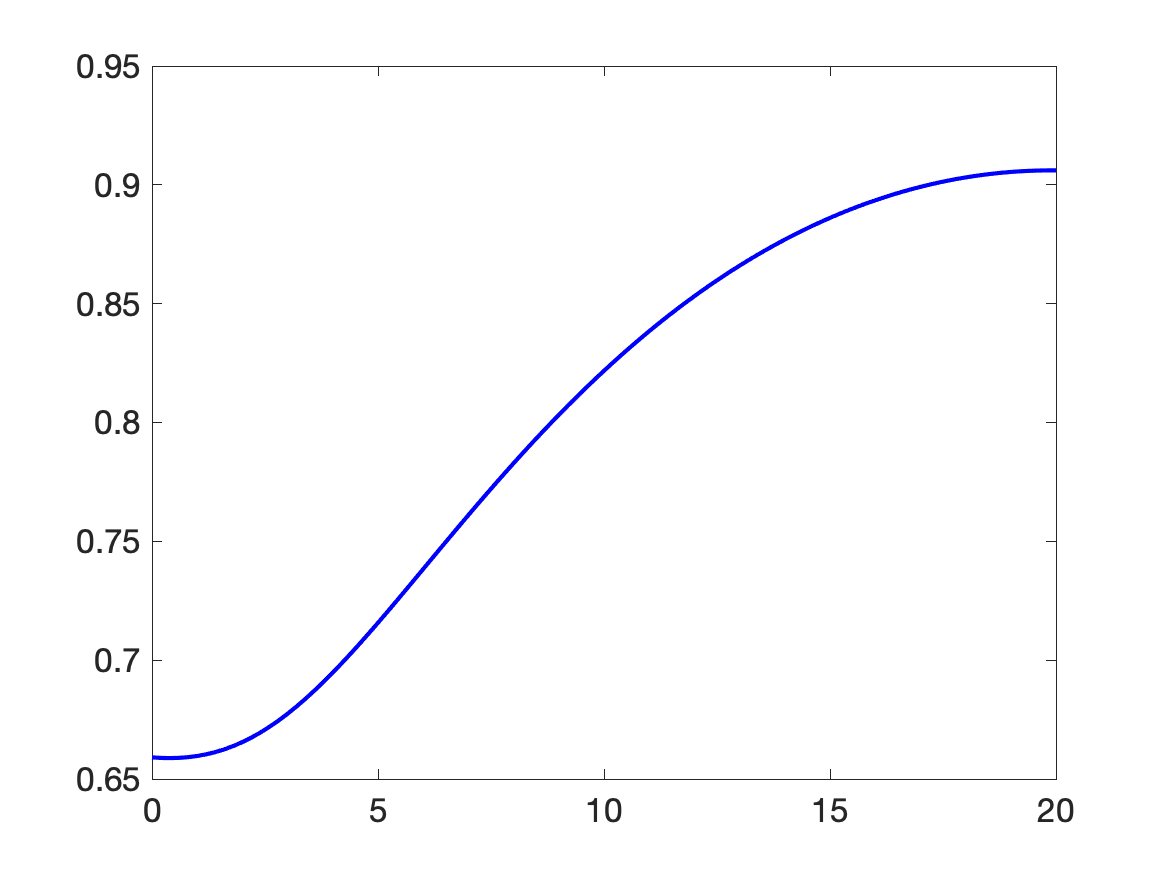}};
\node at ([xshift=3mm]imgB.west) {\tiny $h$};
\node at ([yshift=1mm]imgB.south) {\tiny $x$};
\end{tikzpicture}
\caption{$\tau=0.01$}
\end{subfigure}

\caption{Height of the beam cross-section for $p = 0.1\,\mathrm{N/dm}$ and $\varepsilon^p = -0.01$,
shown for $\tau = +\infty$ and $\tau = 0.01$ after 10 time steps.}
\label{beamp4}
\end{figure}

The solutions with $p = 0.1\,\mathrm{N/dm}$ and $\varepsilon^p=-0.01$ are depicted in Figure \ref{beamp4}. We notice that for $\tau=+\infty$ the height of the cross-section has large variations in a narrow space.

\subsection{Case $\varepsilon_i^p = 0$ and $\kappa_i^p =\kappa^p$ for all $i$}
We consider the case in the prestrain is null and the precurvature is constant for all $i$.
The beam is subjected to a bending moment
$M(x)$.
The expression of the mean compliance changes at each iteration.
At the first iteration, it is given by
\begin{equation}\label{C12}
C_1(h_1)
= \int_{0}^{\ell}
\begin{multlined}[t]
\frac{4\left(E\,\kappa^p h_{0}^3+3M\right)^2}{3E h_{1}^3}
-\frac{\kappa^p\left(2E\,\kappa^p h_{0}^3-E\,\kappa^p h_{1}^3+6M\right)}{3} \\
-\frac{2h_{0}^2\kappa^p\left(E\,\kappa^p h_{0}^3+3M\right)}{h_{1}^2}
+\frac{E h_{0}^4(\kappa^p)^2}{h_{1}}
\, dx .
\end{multlined}
\end{equation}

Introducing the dimensionless quantities
\[
\hbar(x) := \frac{h_1(x)}{h_0(x)}
\qquad \text{and} \qquad
\mu(x) := \frac{M(x)}{E h_0^3(x)\kappa^p(x)},
\]
the mean compliance \eqref{C12} can be rewritten as
\[
C_1(h_1)
= \int_{0}^{\ell}
E h_0^3(x)\bigl(\kappa^p(x)\bigr)^2
\, g\bigl(\mu(x),\hbar(x)\bigr)
\, dx,
\]
where the function $g$ is defined by
\[
g(\mu,\hbar)
:= \frac{1}{\hbar}
-\frac{-\hbar^3+6\mu+2}{3}
-\frac{2(3\mu+1)}{\hbar^2}
+\frac{4(3\mu+1)^2}{3\hbar^3}.
\]

The second derivative of $g$ with respect to $\hbar$ reads
\[
\begin{aligned}
\frac{\partial^2 g}{\partial \hbar^2}(\mu,\hbar)
&= \frac{2\left(
	72\mu^2-18\mu \hbar+48\mu+\hbar^6+\hbar^2-6\hbar+8
	\right)}{\hbar^5}\\
	&= \frac{2}{\hbar^5}\Big(72(\mu + \frac{8-3\hbar}{24})^2 + \frac{\hbar^2(8\hbar^4-1)}{8}\Big).
\end{aligned}
\]
From this expression, since $\hbar \geq 1$ we deduce that
$
\frac{\partial^2 g}{\partial \hbar^2}(\mu,\hbar) > 0$.
Hence, at least $C_1$ is convex.

\subsubsection{Constant bending moment}
The beam is subjected to a constant bending moment
$M = 20\,\mathrm{N\,dm}$.
Independently of the sign of $\kappa^p$ and of the value of $\tau$ (possibly equal to $+\infty$),
the solution of problem~\eqref{prob_disc1_cor} is always the same: the beam grows by keeping the cross-section height constant along the length of the beam. The results are identical with those reported in Figure~\ref{beam2}.

\subsubsection{Parabolic bending moment}

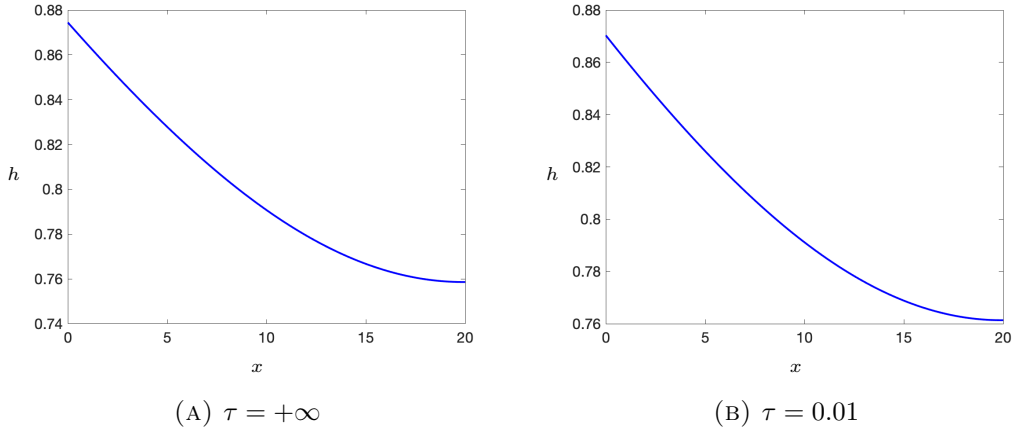
\begin{figure}[!hbt]
\centering

% ---- left panel ----
\begin{subfigure}{0.45\textwidth}
\centering
\begin{tikzpicture}
\node (imgA) {\includegraphics[width=\linewidth]{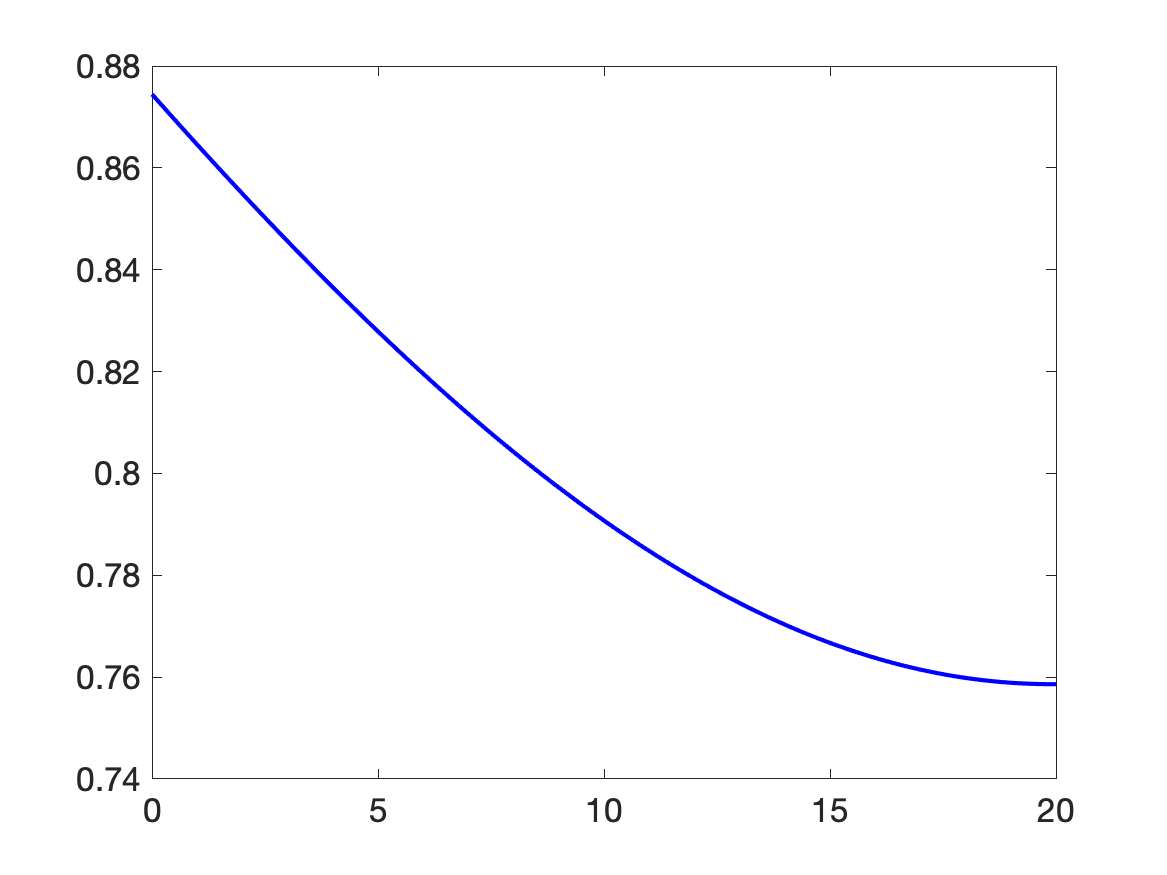}};
% y-label for this panel (change -6mm if needed)
\node at ([xshift=3mm]imgA.west) {\tiny $h$};
% x-label for this panel (change -6mm if needed)
\node at ([yshift=1mm]imgA.south) {\tiny $x$};
\end{tikzpicture}
\caption{$\tau=+\infty$}
\end{subfigure}
\hspace{0.1cm}
% ---- right panel ----
\begin{subfigure}{0.45\textwidth}
\centering
\begin{tikzpicture}
\node (imgB) {\includegraphics[width=\linewidth]{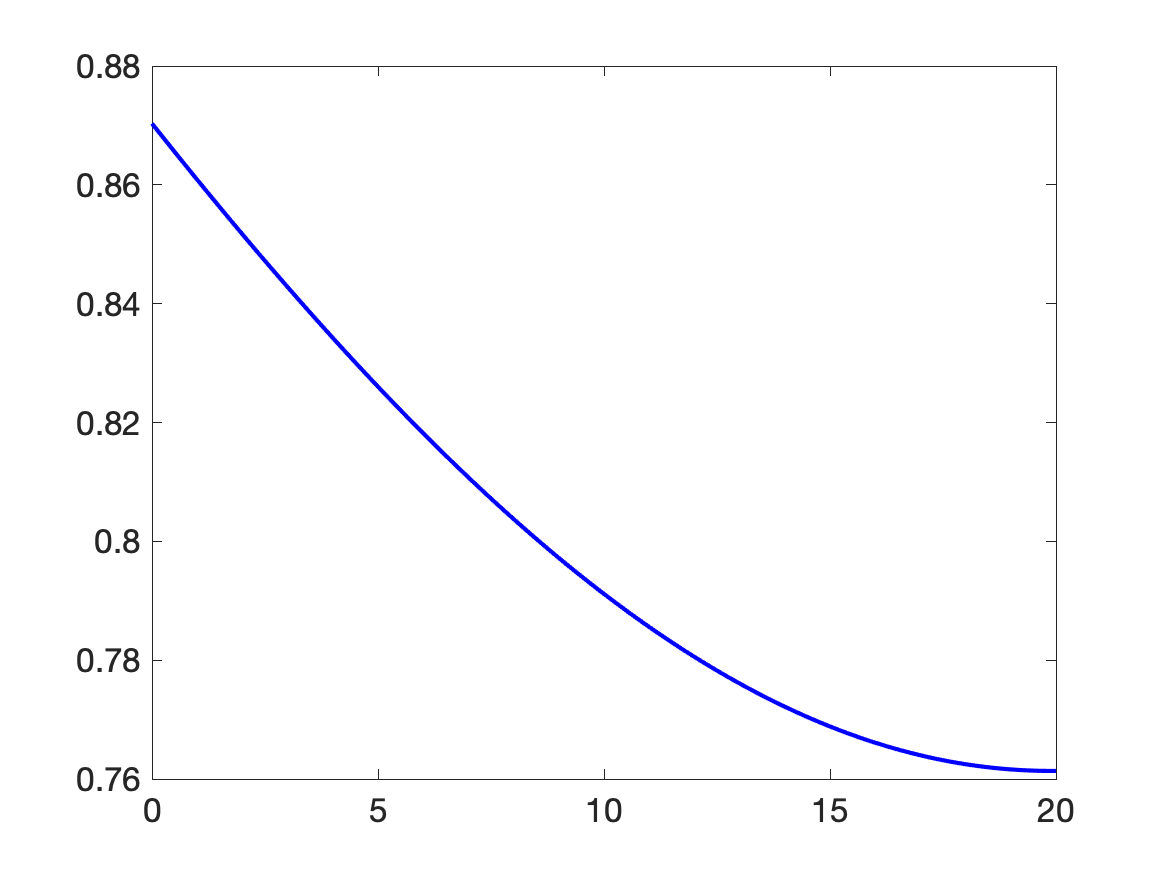}};
\node at ([xshift=3mm]imgB.west) {\tiny $h$};
\node at ([yshift=1mm]imgB.south) {\tiny $x$};
\end{tikzpicture}
\caption{$\tau=0.01$}
\end{subfigure}

\caption{Parabolic bending moment: height of the beam cross-section for $p = 0.1\,\mathrm{N/dm}$, $\varepsilon^p=0$, $\kappa^p = 0.05$,
shown for $\tau = +\infty$ and $\tau = 0.01$ after 10 time steps.}
\label{hk1}
\end{figure}

We consider the case in which the beam is subjected to a uniform load of intensity $p$, inducing the bending moment given by \eqref{M0}. The results obtained for $\kappa^p = 0.05$ and $p = 0.1\,\mathrm{N/dm}$  are depicted in Figure \ref{hk1}.

\begin{figure}[!hbt]
\centering

% ---- left panel ----
\begin{subfigure}{0.45\textwidth}
\centering
\begin{tikzpicture}
\node (imgA) {\includegraphics[width=\linewidth]{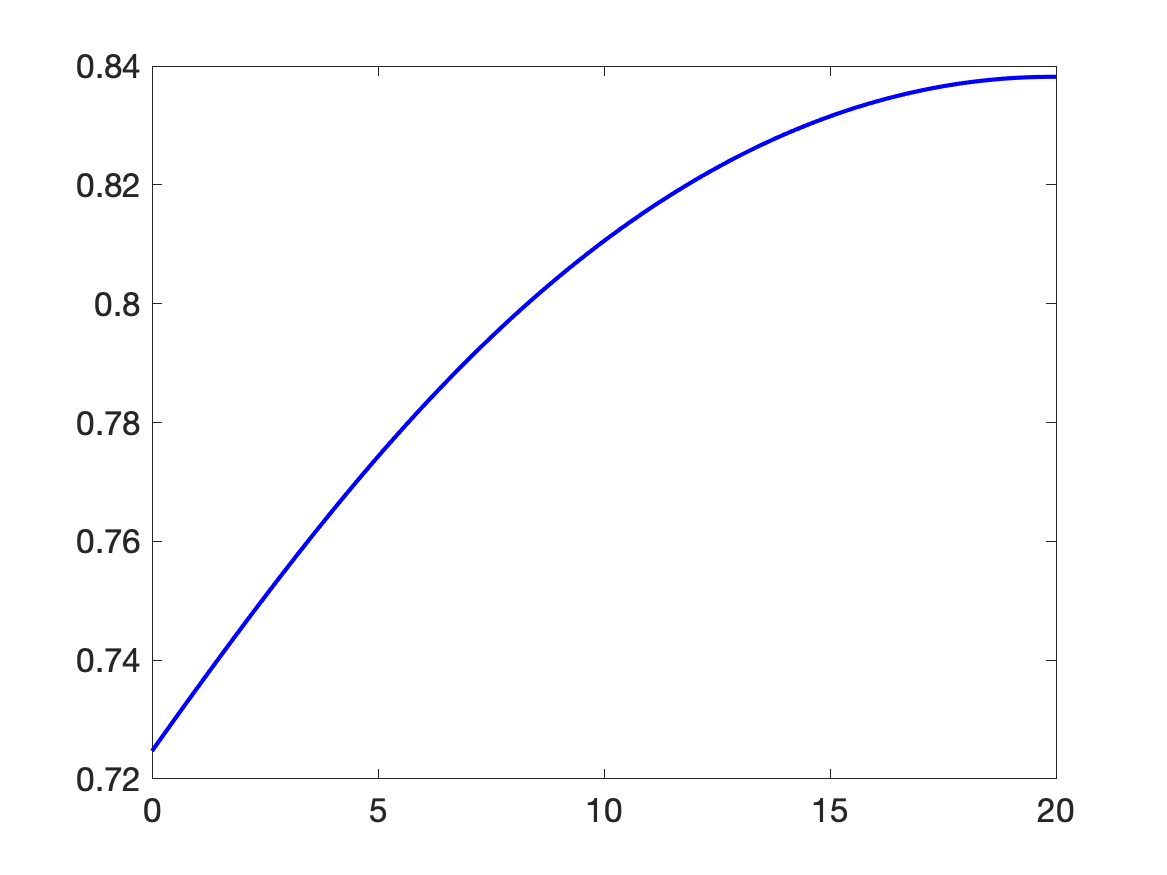}};
% y-label for this panel (change -6mm if needed)
\node at ([xshift=3mm]imgA.west) {\tiny $h$};
% x-label for this panel (change -6mm if needed)
\node at ([yshift=1mm]imgA.south) {\tiny $x$};
\end{tikzpicture}
\caption{$\tau=+\infty$}
\end{subfigure}
\hspace{0.1cm}
% ---- right panel ----
\begin{subfigure}{0.45\textwidth}
\centering
\begin{tikzpicture}
\node (imgB) {\includegraphics[width=\linewidth]{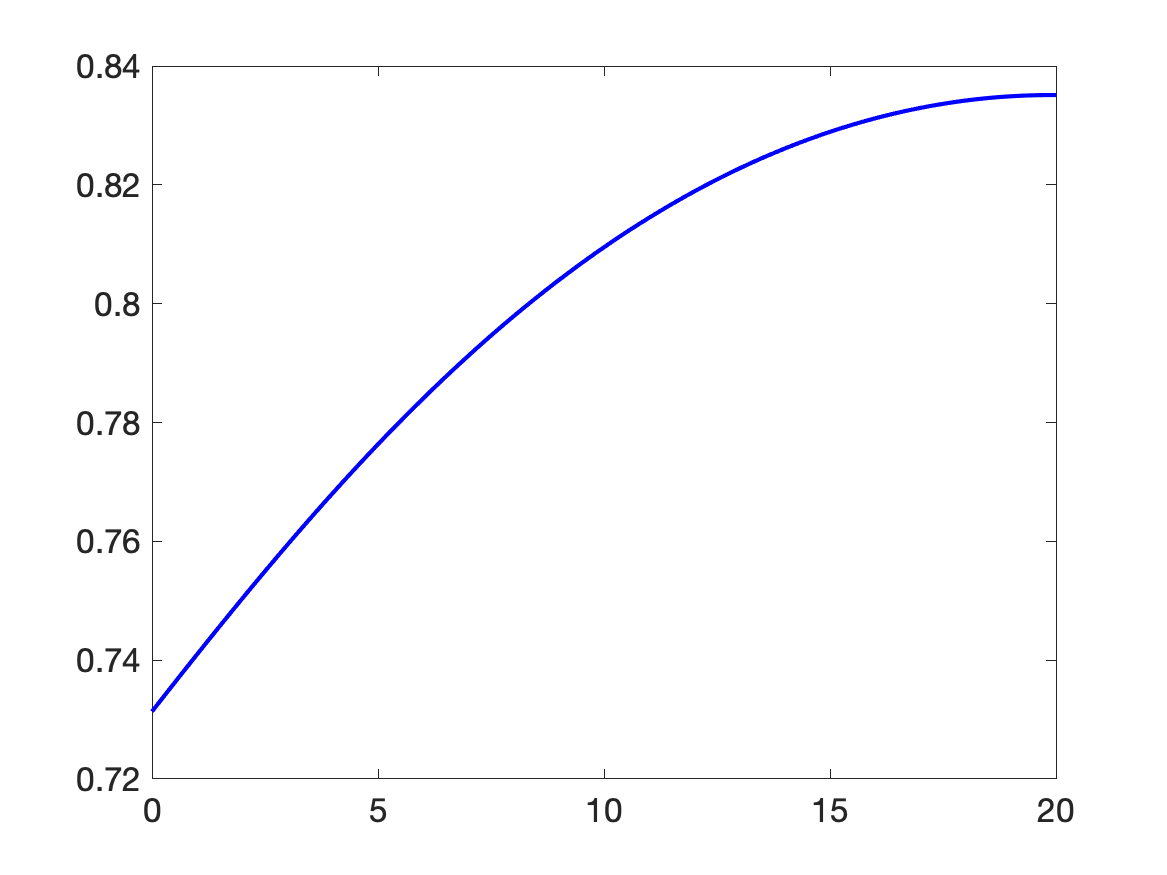}};
\node at ([xshift=3mm]imgB.west) {\tiny $h$};
\node at ([yshift=1mm]imgB.south) {\tiny $x$};
\end{tikzpicture}
\caption{$\tau=0.01$}
\end{subfigure}

\caption{Parabolic bending moment: height of the beam cross-section for $p = 0.1\,\mathrm{N/dm}$, $\varepsilon^p=0$, $\kappa^p = -0.05$,
shown for $\tau = +\infty$ and $\tau = 0.01$ after 10 time steps.}
\label{hk2}
\end{figure}
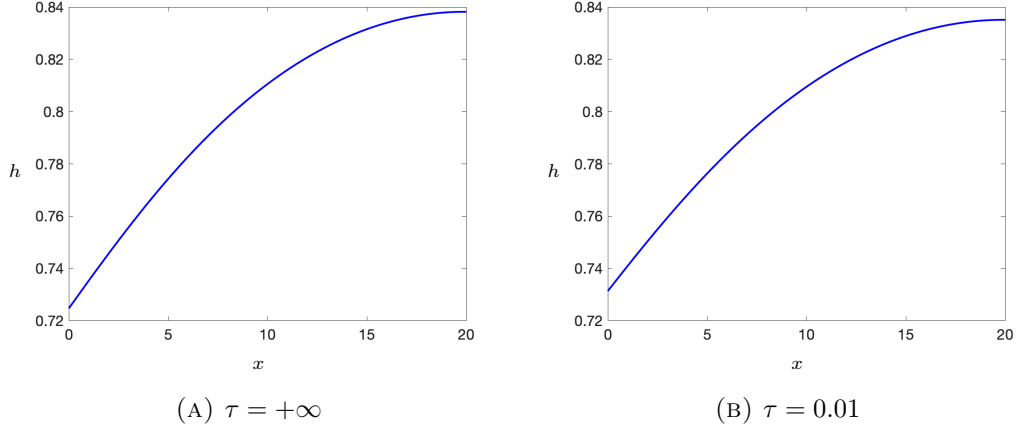

The results obtained for $\kappa^p = -0.05$ and $p = 0.1\,\mathrm{N/dm}$  are depicted in Figure \ref{hk2}.

\subsection{General remarks}
The following generalizations can be readily handled:
\begin{enumerate}
    \item The structure considered is statically determinate; therefore, the bending moment does not depend on the cross-sectional height. Nevertheless, the procedure described can also handle, albeit with more tedious calculations, structures that are statically indeterminate.
    \item The boundary and loading conditions may vary at each time step.
    \item As observed in Remark \ref{rem3}, in problem \eqref{prob_disc1_cor_less} the constraint
    $$ \int_0^\ell h_i(x)\, dx = m_i$$
    can be replaced by
    $$ \int_0^\ell h_i(x)\, dx \le m_i.$$
In this case, $m_i$ represents the maximum mass that the beam is allowed to have at step~$i$.
    With the inequality constraint, the beam may decide whether or not to absorb the material supplied to it.
\item Problem \eqref{prob_disc1_cor_less} does not allow ablation. To include this possibility,
it suffices to remove the constraint
$$
h_i(x) \ge h_{i-1}(x), \quad x \in [0,\ell],
$$
from problem \eqref{prob_disc1_cor_less}.
\end{enumerate}

\section{Continuous-time formulation}

The aim of this section is to investigate whether a time-continuous formulation can be derived from the time-discrete problem \eqref{prob_disc1_cor}.
The calculations that follow are purely formal. We do not attempt to provide a rigorous justification; rather, our goal is to offer a heuristic explanation.

Let us denote the density of the compliance $C_i$ by $c(\cdot;\varepsilon^p_i,\kappa^p_i)$, so to write
$$
C_i(h_i)=\int_0^\ell c(h_i;\varepsilon^p_i,\kappa^p_i) \, dx.
$$
With this notation, problem \eqref{prob_disc1_cor}  writes as
\begin{equation}\label{prob_i}
\left\{
\begin{aligned}
\min_{h_i} \quad
& \int_0^\ell c(h_i;\varepsilon^p_i,\kappa^p_i) \, dx+ \frac{1}{2\tau} \int_0^\ell |h_i - h_{i-1}|^2 \, dx, \\
& \int_0^\ell h_i(x)\, dx = m_i, \\
& h_i(x) \ge h_{i-1}(x), \quad x \in [0,\ell].
\end{aligned}
\right.
\end{equation}
For every $i$, let $h_i$ be a minimizer of problem \eqref{prob_i}
and let
$$
\mathcal{A}_i:=\Big\{\varphi:(0,\ell)\to\R : \int_0^\ell \varphi\,dx=0  \mbox{ and } \varphi=0 \mbox{ on }\{x:h_i(x)=h_{i-1}(x)\}
\Big\}.
$$
For $\varphi_i\in\mathcal{A}_i$ and for $s\in \R$ sufficiently small it follows that
$$
\int_0^\ell (h_i(x)+s\varphi_i(x))\, dx = m_i  \mbox{ and } h_i(x)+s\varphi_i(x) \ge h_{i-1}(x), \quad x \in [0,\ell],
$$
hence, by taking variations in problem \eqref{prob_i} we deduce that $h_i$ satisfies the following problem:
\begin{equation}\label{prob_i_wf}
\left\{
\begin{aligned}
& \int_0^\ell \big(c'(h_i;\varepsilon^p_i,\kappa^p_i) + \frac{1}{\tau} (h_i - h_{i-1})\big)\varphi_i \, dx=0\qquad \forall\varphi_i\in\mathcal{A}_i,\\
& \int_0^\ell h_i(x)\, dx = m_i, \\
& h_i(x) \ge h_{i-1}(x), \quad x \in [0,\ell],
\end{aligned}
\right.
\end{equation}
where $c'$ denotes the derivative of $c$ with respect to the first variable.

Let $t\mapsto m(t)$ be the function that prescribes the mass of the beam at time $t$ and such that
$$m_i=m(i\tau)\qquad \forall i.$$
We also denote by
$$
m^\tau(t):=m(\lfloor t/\tau \rfloor \tau)
$$
where, for $a\in\R$, $\lfloor a \rfloor$ denotes the largest integer smaller than $a$.

\begin{center}
\begin{tikzpicture}[>=Stealth]

% Axes
\draw[-latex] (0,0) -- (7,0) node[right] {$t$};
\draw[-latex] (0,0) -- (0,4.5) node[above] {$m$};

% Visual spacing for tau
\def\tauv{1.8}

% x-axis ticks: 1tau, 2tau, 3tau
\foreach \k in {1,2,3}{
    \draw (\k*\tauv,0) -- (\k*\tauv,-0.1)
        node[below] {$\k\tau$};
}

% Increasing smooth function m(t)
\draw[thick,blue,domain=0:6.5,samples=200]
    plot (\x,{1 + 0.6*exp(0.25*\x)})
    node[right] {$m(t)$};

% Mark grid values m_k = m(k tau)
\foreach \k in {0,1,2,3}{
    \pgfmathsetmacro{\tk}{\k*\tauv}
    \pgfmathsetmacro{\mk}{1 + 0.6*exp(0.25*\tk)}
    \fill[blue] (\tk,\mk) circle (1.5pt);
}

% Step function m^tau(t)
\foreach \k in {0,1,2,3}{
    \pgfmathsetmacro{\tleft}{\k*\tauv}
    \pgfmathsetmacro{\tright}{(\k+1)*\tauv}
    \pgfmathsetmacro{\mleft}{1 + 0.6*exp(0.25*\tleft)}
    \pgfmathsetmacro{\mright}{1 + 0.6*exp(0.25*\tright)}

    % horizontal segment
    \draw[thick,red] (\tleft,\mleft) -- (\tright,\mleft);
    % vertical jump
    %\draw[thick,red] (\tright,\mleft) -- (\tright,\mright);

    % closed-left / open-right
    \fill[red] (\tleft,\mleft) circle (1.5pt) node[above left] {$m_{\k}$};
    \draw[red,fill=white] (\tright,\mleft) circle (1.5pt);
}

\node[red] at (5.5,1.8) {$m^\tau(t)$};

\end{tikzpicture}
\end{center}
For later use, note that
$$
\mbox{if }t\in[i\tau,(i+1)\tau)\quad\Rightarrow\quad m^\tau(t)=m_i\mbox{ and } m^\tau(t-\tau)=m_{i-1}
$$
and, since $t-\tau\le \lfloor t/\tau \rfloor \tau\le t$,  that
$$
\lim_{\tau\to 0} m^\tau(t)=\lim_{\tau\to 0}  m(\lfloor t/\tau \rfloor \tau)= m(t)\quad \forall t,
$$
at least if $m$ is continuous.

Similarly, let
$$
h^\tau(x,t):=h_i(x),\qquad \varphi^\tau(x,t):=\varphi_i(x)\qquad \mbox{if }t\in[i\tau,(i+1)\tau),
$$
and
$$
(\varepsilon^p)^\tau(x,t):=\varepsilon^p_i(x),\qquad (\kappa^p)^\tau(x,t):=\kappa^p_i(x)\qquad \mbox{if }t\in[i\tau,(i+1)\tau),
$$

With this notation problem \eqref{prob_i_wf} writes as

\begin{equation}\label{prob_i_wf2}
\left\{
\begin{aligned}
& \int_0^\ell [c'(h^\tau(x,t);(\varepsilon^p)^\tau(x,t),(\kappa^p)^\tau(x,t)) + \frac{h^\tau(x,t)-h^\tau(x,t-\tau)}{\tau}]\varphi^\tau(x,t) \, dx=0\\
& \hspace{11cm} \forall\varphi^\tau(\cdot,t)\in\mathcal{A}_{\lfloor t/\tau \rfloor},\\
& \int_0^\ell h^\tau(x,t)\, dx = m^\tau(t), \\
&  h^\tau(x,t)-h^\tau(x,t-\tau)\ge 0, \quad x \in [0,\ell].
\end{aligned}
\right.
\end{equation}
Assume that the following limits exist
$$
\lim_{\tau\to 0} h^\tau(x,t)= h(x,t)\quad \mbox{and}\quad  \lim_{\tau\to 0}\frac{h^\tau(x,t)-h^\tau(x,t-\tau)}{\tau}=\frac{\partial h}{\partial t}(x,t) \qquad\forall x,t
$$
and
$$
\lim_{\tau\to 0} \varphi^\tau(x,t)= \varphi(x,t)\quad \lim_{\tau\to 0} (\varepsilon^p)^\tau(x,t)=\varepsilon^p(x,t)
\quad  \lim_{\tau\to 0} (\kappa^p)^\tau(x,t)=\kappa^p(x,t)\quad\forall x,t,
$$
for some functions $h$, which are called minimizing movements, $\varphi$, $\varepsilon^p$, and $\kappa^p$.
By taking the limit for $\tau\to 0$ in problem \eqref{prob_i_wf2} we deduce
\begin{equation}\label{prob_i_wf3}
\left\{
\begin{aligned}
& \int_0^\ell [c'(h(x,t);\varepsilon^p(x,t),\kappa^p(x,t)) + \frac{\partial h}{\partial t}(x,t) ]\varphi(x,t) \, dx=0\quad \forall\varphi\in\mathcal{A},\\
& \int_0^\ell h(x,t)\, dx = m(t), \\
& \frac{\partial h}{\partial t}(x,t) \ge 0,
\end{aligned}
\right.
\end{equation}
where
$$
\begin{aligned}
\mathcal{A}:=\Big\{\varphi:(0,\ell)\times (0,+\infty)\to\R : & \int_0^\ell \varphi(x,t)\,dx=0 \\
& \mbox{ and } \varphi=0 \mbox{ on }\{(x,t):  \frac{\partial h}{\partial t}(x,t)=0\}
\Big\}.
\end{aligned}
$$
Up to a rigorous justification, \eqref{prob_i_wf3}
can be interpreted as the time-continuous problem, in a weak form,
obtained as a limit of the time-discrete problem~\eqref{prob_disc1_cor_less}.

\section{Closed form analytical solution for the baseline case}\label{app1}

We here analytically study the problem posed in Section \ref{section_00}:
\begin{equation}\label{prob_app}
\left\{
\begin{aligned}
\min_{h_i} \quad &  \int_{0}^{\ell} \frac{12\, M(x)^2}{E\,h_i(x)^3} \, dx, \\
\quad
& \int_0^\ell h_i(x)\, dx = m_i, \\
& h_i(x) \ge h_{i-1}(x), \quad x \in [0,\ell],
\end{aligned}
\right.
\end{equation}
where the bending moment is given by \eqref{M0}.

To find the solution we make use of the Karush–Kuhn–Tucker (KKT) conditions, which can be  partly deduced from the
Lagrangian
\[
\begin{array}{lll}
\mathcal{L}(h_i, \mu_i, \lambda_i)
= &\displaystyle \int_0^\ell \frac{12\, M(x)^2}{E\,h_i(x)^3} \, dx
+ \int_0^\ell \mu_i(x) \, (h_{i-1}(x) - h_i(x)) \, dx\\[2ex]
&\displaystyle \hspace{4cm} + \lambda_i \left( \int_0^\ell h_i(x) \, dx - m_i \right),
\end{array}\eqno(i)
\]
where $\lambda_i\in\R$ and $\mu_i:[0,\ell]\to [0,+\infty)$ are the Lagrange multipliers.

The KKT conditions write as:
\begin{align}
-\frac{36\, M(x)^2}{E\,h_i(x)^4} - \mu_i(x) + \lambda_i &= 0, \quad \text{on } [0, \ell], \tag{\(ii\)}\\
h_{i-1} - h_i(x) &\le 0, \quad \text{on } [0, \ell], \tag{\(iii\)}\\
\int_0^\ell h_i(x) \, dx - m_i &= 0, \tag{\(iv \)}\\
\mu_i(x) &\ge 0, \quad \text{on } [0, \ell], \tag{\(v\)}\\
\mu_i(x)\,(h_{i-1} - h_i(x)) &= 0, \quad \text{on } [0, \ell]. \tag{\(vi\)}
\end{align}

Equations  \((ii)\) and  \((iv)\) are the `derivatives' of the Lagrangian with respect to $h_i$ and $\lambda_i$, respectively,
 \((iii)\) is simply the third equation of problem \eqref{prob_disc1_3},  \((v)\) assures the positivity of the multiplier $\mu_i$, and   \((vi)\) is the so called complementary condition.
%%%

From  \((vi)\) we deduce that either \(\mu_i(x) = 0\) or \(h_i(x) = h_{i-1}(x)\), on \([0, \ell]\).

%%%%%%%%%%%%%%%%%%%%%%%%%%%%%%%%%

\begin{itemize}
\item Case \(\mu_i(x) = 0.\)

\noindent
From (\(ii\)), we find that
$$
\lambda_i =\frac{36\, M(x)^2}{E\,h_i(x)^4} \eqno(vii)
$$
that implies
\[
h_i(x) = \left( \frac{36\, M(x)^2}{E\,\lambda_i }  \right)^{1/4}. \eqno(viii)
\]
This identity holds where  ($iii$)  is satisfied, that is on
$$
I_i=\left\{x\in [0,\ell]:\left( \frac{36\, M(x)^2}{E\,\lambda_i }  \right)^{1/4}\ge h_{i-1}(x)\right\}.\eqno(ix)
$$

\item Case \(h_i(x) = h_{i-1}(x)\).

\noindent
Equation ($ii$) leads to
\[
\mu_i(x)= -\frac{36\,M(x)^2}{E\,h_{i-1}(x)^4}   + \lambda_i,
\]
and ($v$) implies that this solution holds for all $x$ for which
\[
-\frac{36\,M(x)^2}{E\,h_{i-1}(x)^4}   + \lambda_i\ge 0,
\]
that is, for $x\in [0,\ell]\setminus I_i$.
\end{itemize}
Thus, we have  found that
\[
h_i(x) = \left\{
\begin{array}{lll}
\displaystyle \left( \frac{36\, M(x)^2}{E\,\lambda_i }  \right)^{1/4} \quad &{\rm for} \qquad & x \in I_i\\[2ex]
h_{i-1}  \quad &{\rm for} \qquad & x \in [0,\ell]\setminus I_i.
\end{array}
\right. \eqno(x)
\]

With the bending moment given by \eqref{M0}, we deduce that
\[
h_i(x) = \left\{
\begin{array}{lll}
\displaystyle \left( \frac{9\, p^2}{E\,\lambda_i }  \right)^{1/4}(\ell-x) \quad &{\rm for} \qquad & x \in I_i\\[2ex]
h_{i-1}  \quad &{\rm for} \qquad & x \in [0,\ell]\setminus I_i.
\end{array}
\right. \eqno(xi)
\]
From ($xi$) it follows that
$h_i$ is affine on a subset of $(0,\ell)$ and coincides with $h_{i-1}$ on the complementary subset,
as stated in Section \ref{section_00}.

Finally, the multiplier $\lambda_i$ can be deduce by imposing the mass constraint \((iv)\).
For instance, for $i=1$, recalling that $h_0$ is constant, we deduce that
$$
I_1=\left\{x\in [0,\ell]:\left( \frac{9\, p^2}{E\,\lambda_1 }  \right)^{1/4}(\ell-x)\ge h_{0}\right\}
=\{x\in [0,\ell]: x\le \hat x\},\eqno(xii)
$$
where we have set
\[
\hat{x} := \ell - h_0 \left( \frac{E\,\lambda_1 }{9\, p^2}  \right)^{1/4}. \eqno(xiii)
\]

Using equation \((iv)\), the mass constraint,
\[
\int_0^{\hat{x}} \left( \frac{9\, p^2}{E\,\lambda_1 }  \right)^{1/4}(\ell-x) \, dx + (\ell - \hat{x})h_0 = m_i
\]
that implies
\[
\left( \frac{9\, p^2}{E\,\lambda_1}  \right)^{1/4} = \frac{ m_1\pm \sqrt{ m_1^2-h_0^2\ell^2}}{\ell^2}
\eqno(xiv)
\]
Imposing that $\hat x\ge 0$, we find that
\[
\left( \frac{9\, p^2}{E\,\lambda_1}  \right)^{1/4} = \frac{ m_1+ \sqrt{ m_1^2-h_0^2\ell^2}}{\ell^2},
\eqno(xv)
\]
and therefore
\[
h_1(x) = \left\{
\begin{array}{lll}
\displaystyle h_0\frac{ m_1+ \sqrt{ m_1^2-m_0^2}}{m_0}\frac{\ell-x}\ell \quad &{\rm for} \qquad & 0\le x \le\hat x,\\[2ex]
h_{0}  \quad &{\rm for} \qquad & \hat x < x \le \ell,
\end{array}
\right. \eqno(xvi)
\]
where $m_0=h_0\ell$ and
$$
\hat{x} = \Big(1 -\frac{m_0}{ m_1+ \sqrt{ m_1^2-m_0^2}}\Big)\ell. \eqno(xvii)
$$

\section{Conclusions}

In this work we have proposed a variational framework for surface growth driven by an optimality criterion. Growth is described as a discrete-time process in which material is progressively deposited on the boundary, and the evolving geometry at each step is determined as the solution of a constrained minimization problem. The objective functional is chosen so as to capture relevant features of the biological or physical system under consideration, while the constraints enforce equilibrium, mass balance, and irreversibility of deposition. In this way, rather than prescribing a growth velocity or a surface mass flux, the geometric evolution is obtained from an optimization principle; in the examples presented here, the objective functional is the compliance.

Although the analysis has been carried out in the simplified setting of a linearly elastic cantilever beam with variable cross-sectional height, the model captures several general aspects of accretive growth. In particular, growth-induced prestrain and precurvature generate residual stresses that significantly affect the optimal redistribution of material, even in statically determinate structures. The results further indicate that the convexity properties of the compliance functional play a decisive role in determining the qualitative behavior of the solution: when convexity is preserved, the growth profiles are smooth and stable, whereas loss of convexity may lead to localization, non-uniqueness, and numerical instabilities.

To mitigate these effects, we introduced a regularized incremental formulation inspired by the theory of minimizing movements. The additional penalization term favors configurations close to the previous one and restores well-posedness in regimes where the original functional is non-convex. This regularized scheme provides a natural interpretation of growth as a variational evolution process and establishes a formal connection between the discrete-time formulation and a possible continuous-time limit, which has been derived at a heuristic level.

More generally, the proposed framework shows that optimality principles can be incorporated explicitly into continuum descriptions of surface growth. Even within a minimal mechanical setting, this perspective highlights how geometry, residual stress, and functional performance interact in shaping structural evolution. The formulation can be extended to more complex geometries, statically indeterminate systems, nonlinear constitutive evolutions, and alternative objective functionals.

Surface growth can thus be interpreted not merely as mass accumulation, but as an adaptive process in which the structure evolves toward configurations that optimize its mechanical response under prescribed constraints. This viewpoint opens promising directions for both the modeling of biological growth and the design of engineering systems capable of morphology-driven adaptation.

\medskip

\noindent
{\bf Acknowledgments.} We gratefully acknowledge the support of the MIT-UNIPI Seed Fund.  R.P.  and  M.P.S. thankfully acknowledges the support of the Italian National Group of Mathematical Physics (GNFM-INdAM).

\color{black}


\begin{thebibliography}{}

	\bibitem[Akerson et al.(2022)]{Akerson2022}
	Akerson, A., Bourdin, B., Bhattacharya, K. (2022).
	\newblock Optimal design of responsive structures.
	\newblock {\em Structural and Multidisciplinary Optimization}, 65.
	\newblock doi:10.1007/s00158-022-03202-0.

	\bibitem[Allaire et al.(2021)]{Allaire2021}
	Allaire, G., Dapogny, C., Jouve, F. (2021).
	\newblock Shape and topology optimization.
	\newblock In: Bonito, A., Nochetto, R.H. (eds.),
	{\em Handbook of Numerical Analysis}, Vol.~22, 1--132.
	\newblock Elsevier.
	\newblock doi:10.1016/bs.hna.2020.10.004.

	\bibitem[Ambrosi and Mollica(2002)]{Ambrosi2002}
Ambrosi, D., Mollica, F. (2002).
\newblock On the mechanics of a growing tumor.
\newblock {\em International Journal of Engineering Science}, 40(12), 1297--1316.

	\bibitem[Ambrosi and Guana(2007)]{Ambrosi2007}
	Ambrosi, D., Guana, F. (2007).
	\newblock Stress-modulated growth.
	\newblock {\em Mathematics and Mechanics of Solids}, 12(3), 319--342.
	\newblock doi:10.1177/1081286505059739.

	\bibitem[Ambrosio et al.(2005)]{Ambrosio2005}
	Ambrosio, L., Gigli, N., Savaré, G. (2005).
	\newblock {\em Gradient Flows in Metric Spaces and in the Space of Probability Measures}.
	\newblock Birkhäuser, Basel.
	\newblock doi:10.1007/978-3-7643-8722-8.

	\bibitem[Andrini et al.(2022)]{Andrini2022}
Andrini, D., Noselli, G., Lucantonio, A. (2022).
\newblock Optimal design of planar shapes with active materials.
\newblock {\em Proceedings of the Royal Society A: Mathematical, Physical and Engineering Sciences}, 478, 20220256.

	\bibitem[Bendsøe and Sigmund(2003)]{Bendsoe2003}
	Bendsøe, M.P., Sigmund, O. (2003).
	\newblock {\em Topology Optimization: Theory, Methods, and Applications}.
	\newblock Springer, Berlin.
	\newblock doi:10.1007/978-3-662-05086-6.

	\bibitem[Braides(2014)]{Braides2014}
	Braides, A. (2014).
	\newblock {\em Local Minimization, Variational Evolution and $\Gamma$-Convergence}.
	\newblock Lecture Notes in Mathematics.
	\newblock Springer, Cham.
	\newblock doi:10.1007/978-3-319-01982-6.

	\bibitem[Chen and Hoger(2000)]{Chen2000}
	Chen, Y.-C., Hoger, A. (2000).
	\newblock Constitutive functions of elastic materials in finite growth and deformation.
	\newblock {\em Journal of Elasticity}, 59, 175--193.
	\newblock doi:10.1023/A:1011061400438.

	\bibitem[De Giorgi(1993)]{DeGiorgi1993}
	De Giorgi, E. (1993).
	\newblock New problems on minimizing movements.
	\newblock In: {\em Boundary Value Problems for PDEs and Applications}, 81--98.
	\newblock Masson.

	\bibitem[Di Carlo and Quiligotti(2002)]{DiCarloQuiligotti2002}
	Di Carlo, A., Quiligotti, S. (2002).
	\newblock Growth and balance.
	\newblock {\em Mechanics Research Communications}, 29, 449--456.


\bibitem[Epstein and Maugin(2000)]{Epstein2000}
	Epstein, M., Maugin, G.A. (2000).
	\newblock Thermomechanics of volumetric growth in uniform bodies.
	\newblock {\em International Journal of Plasticity}, 16, 951–978.
	\newblock doi:10.1016/S0749-6419(99)00081-9.

		\bibitem[Erlich and Zurlo(2024)]{Erlich2024}
Erlich, A., Zurlo, G. (2024).
\newblock Incompatibility-driven growth and size control during development.
\newblock {\em Journal of the Mechanics and Physics of Solids}, 188, 105660.

\bibitem[Erlich and Zurlo(2025)]{Erlich2025}
Erlich, A., Zurlo, G. (2025).
\newblock The geometric nature of homeostatic stress in biological growth.
\newblock {\em Journal of the Mechanics and Physics of Solids}, 201, 106155.


	\bibitem[Goriely(2017)]{Goriely2017}
	Goriely, A. (2017).
	\newblock {\em The Mathematics and Mechanics of Biological Growth}.
	\newblock Springer, New York.
	\newblock doi:10.1007/978-1-4939-2106-3.

	\bibitem[Goriely and Ben Amar(2005)]{Goriely2005}
	Goriely, A., Ben Amar, M. (2005).
	\newblock Differential growth and instability in elastic shells.
	\newblock {\em Journal of the Mechanics and Physics of Solids}, 53, 2284--2319.
	\newblock doi:10.1016/j.jmps.2005.04.004.

	\bibitem[Grillo et al.(2025)]{Grillo2025}
	Grillo, A., Pastore, A., Di~Stefano, S. (2025).
	\newblock An Approach to Growth Mechanics Based on the Analytical Mechanics of Nonholonomic Systems.
	\newblock {\em Journal of Elasticity}, 157, 3.
	\newblock doi:10.1007/s10659-024-10092-7.

	\bibitem[Hoger(1986)]{Hoger1986}
	Hoger, A. (1986).
	\newblock On the determination of residual stress in an elastic body.
	\newblock {\em Journal of Elasticity}, 16, 303--324.
	\newblock doi:10.1007/BF00040812.

	\bibitem[Johnson(2019)]{Johnson2019}
	Johnson, A. T. (2019).
	\newblock {\em Biology for Engineers}.
	\newblock CRC Press.

	\bibitem[Liu et al.(2025)]{Liu2025}
	Liu, J., Costello, J. H., Kanso, E. (2025).
	\newblock Flow physics of nutrient transport drives functional design of ciliates.
	\newblock {\em Nature Communications}, 16, 4154.

	\bibitem[Mattheck(1989)]{Mattheck1989}
	Mattheck, C. (1989).
	\newblock Engineering components grow like trees.
	\newblock Report KfK 4648.
	\newblock Kernforschungszentrum Karlsruhe GmbH.

	\bibitem[Moulton et al.(2013)]{Moulton2013}
	Moulton, D.E., Lessinnes, T., Goriely, A. (2013).
	\newblock Morphoelastic rods. Part I: A single growing elastic rod.
	\newblock {\em Journal of the Mechanics and Physics of Solids}, 61, 398--427.
	\newblock doi:10.1016/j.jmps.2012.09.017.

	\bibitem[Naghibzadeh et al.(2021)]{Naghibzadeh2021}
	Naghibzadeh, S.K., Walkington, N., Dayal, K. (2021).
	\newblock Surface growth in deformable solids using an Eulerian formulation.
	\newblock {\em Journal of the Mechanics and Physics of Solids}, 154, 104499.
	\newblock doi:10.1016/j.jmps.2021.104499.

	\bibitem[Nardinocchi et al.(2025)]{Nardinocchi2025}
	Nardinocchi, P., Puntel, E., Zurlo, G. (2025).
	\newblock Collection of the Journal of Elasticity: Mechanics of Growth and Remodeling in Biology.
	\newblock {\em Journal of Elasticity}, 157, 39.
	\newblock doi:10.1007/s10659-025-10131-x.

	\bibitem[Neiferd et al.(2018)]{Neiferd2018}
	Neiferd, D. J., Grandhi, R. V., Deaton, J. D., Beran, P. S. (2018).
	\newblock Level-set topology optimization of thermoelastic structures: a comparison of compliance, strain energy, and stress objectives.
	\newblock {\em 2018 Multidisciplinary Analysis and Optimization Conference}, 1--20.
	\newblock doi:10.2514/6.2018-3577.

	\bibitem[Ortigosa-Martínez et al.(2024)]{OrtigosaMartinez2024}
Ortigosa-Martínez, R., Martínez-Frutos, J., Mora-Corral, C., Pedregal, P., Periago, F. (2024).
\newblock Shape-programming in hyperelasticity through differential growth.
\newblock {\em Applied Mathematics \& Optimization}, 89, 49.

	\bibitem[Renzi et al.(2024)]{Renzi2024}
	Renzi, D., Marfia, S., Tomassetti, G., Zurlo, G. (2024).
	\newblock A discrete model for layered growth.
	\newblock {\em European Journal of Mechanics A/Solids}, 105, 105232.

	\bibitem[Rodriguez et al.(1994)]{Rodriguez1994}
	Rodriguez, E.K., Hoger, A., McCulloch, A.D. (1994).
	\newblock Stress-dependent finite growth in soft elastic tissues.
	\newblock {\em Journal of Biomechanics}, 27, 455--467.
	\newblock doi:10.1016/0021-9290(94)90021-3.

	\bibitem[Skalak et al.(1982)]{Skalak1982}
	Skalak, R., Dasgupta, G., Moss, M., Otten, M., Dullumeijer, P., Vilmann, H. (1982).
	\newblock Analytical description of growth.
	\newblock {\em Journal of Theoretical Biology}, 94, 555--577.
	\newblock doi:10.1016/0022-5193(82)90301-0.

	\bibitem[Sullivan et al.(2017)]{Sullivan2017}
	Sullivan, T. N., Wang, B., Espinosa, H. D., Meyers, M. A. (2017).
	\newblock Extreme lightweight structures: avian feathers and bones.
	\newblock {\em Materials Today}, 20(7).

	\bibitem[Taber(2020)]{Taber2020}
Taber, L. A. (2020).
\newblock {\em Continuum Modeling in Mechanobiology}.
\newblock Springer, Cham, Switzerland.

	\bibitem[Thompson(1942)]{Thompson1942}
	Thompson, D. W. (1942).
	\newblock {\em On Growth and Form}.
	\newblock Cambridge University Press, Cambridge.

	\bibitem[Tomassetti et al.(2016)]{Tomassetti2016}
	Tomassetti, G., Cohen, T., Abeyaratne, R. (2016).
	\newblock Steady accretion of an elastic body on a hard spherical surface and the notion of a four-dimensional reference space.
	\newblock {\em Journal of the Mechanics and Physics of Solids}, 96, 333--352.

	\bibitem[Truskinovsky and Zurlo(2019)]{Truskinovsky2019}
Truskinovsky, L., Zurlo, G. (2019).
\newblock Nonlinear elasticity of incompatible surface growth.
\newblock {\em Physical Review E}, 99, 053001.

	\bibitem[Zhang et al.(2014)]{Zhang2014}
	Zhang, W., Yang, J., Xu, Y., Gao, T. (2014).
	\newblock Topology optimization of thermoelastic structures: mean compliance minimization or elastic strain energy minimization.
	\newblock {\em Structural and Multidisciplinary Optimization}, 49, 417--429.
	\newblock doi:10.1007/s00158-013-0991-9.

	\bibitem[Zurlo and Truskinovsky(2017)]{Zurlo2017}
Zurlo, G., Truskinovsky, L. (2017).
\newblock Printing non-Euclidean solids.
\newblock {\em Physical Review Letters}, 119, 048001.


	\bibitem[Zurlo and Truskinovsky(2018)]{Zurlo2018}
	Zurlo, G., Truskinovsky, L. (2018).
	\newblock Inelastic surface growth.
	\newblock {\em Mechanics Research Communications}, 93, 174--179.


\end{thebibliography}
\end{document}